\documentclass[twocolumn]{aa}

\usepackage{hyperref}
\usepackage{graphicx}
\usepackage{txfonts}
\usepackage{caption}
\usepackage{subcaption}
\usepackage{rotating}
\usepackage{xcolor}
\usepackage[running]{lineno}

\begin{document}
\title{A Timescale-Resolved Analysis of the Breathing Effect in Quasar Broad Line Regions}

\author{C.-Z. Jiang
      \inst{1,2,3}\thanks{E-mail : wy2020jcz@mail.ustc.edu.cn}
      \and
      J.-X. Wang\inst{1,2,4}\thanks{E-mail : jxw@ustc.edu.cn}
      \and
      H. Sou\inst{1,2}
      \and
      W.-K. Ren\inst{1,2}
      }

\institute{Department of Astronomy, University of Science and Technology of China, Hefei, Anhui 230026, People’s Republic of China
\and
School of Astronomy and Space Science, University of Science and Technology of China, Hefei 230026, People’s Republic of China
\and
Department of Physics and Astronomy, University of California, Riverside, 900 University Ave, Riverside CA 92521, USA
\and
College of Physics, Guizhou University, Guiyang, Guizhou, 550025, People’s Republic of China
}

\abstract
% context heading (optional)
{The single-epoch virial method is a fundamental tool for estimating supermassive black hole (SMBH) masses in large samples of AGNs and has been extensively employed in studies of SMBH–galaxy co-evolution across cosmic time. However, since this method is calibrated using reverberation-mapped AGNs, its validity across the entire AGN population remains uncertain.}
% aims heading (mandatory)
{We aim to examine the breathing effect—the variability of emission line widths with continuum luminosity—beyond reverberation-mapped AGNs, to assess the validity and estimate potential systematic uncertainties of single-epoch virial black hole mass estimates.}
% methods heading (mandatory)
{We construct an unprecedentedly large multi-epoch spectroscopic dataset of quasars from SDSS DR16, focusing on four key broad emission lines (H$\alpha$, H$\beta$, $\textrm{Mg\,\textsc{ii}}$, and $\textrm{C\,\textsc{iv}}$). We assess how breathing behavior evolves with the rest-frame time interval between observations.}
% results heading (mandatory)
{We detect no significant breathing signal in H$\alpha$, H$\beta$, or $\textrm{Mg\,\textsc{ii}}$ at any observed timescale. In contrast, $\textrm{C\,\textsc{iv}}$ exhibits a statistically significant anti-breathing trend, most prominent at intermediate timescales. Notably, for H$\beta$, which has shown breathing in previous reverberation-mapped samples, we recover the effect only in the small subset of quasars with clearly detected BLR lags and only during the epochs when such lags are measurable—suggesting that both the lag and breathing signals are intermittent, possibly due to a weak correlation between optical and ionizing continua. These results highlight the complex, variable, and timescale-dependent nature of line profile variability and underscore its implications for single-epoch black hole mass estimates.}
% conclusions heading (optional), leave it empty if necessary
{}
\maketitle

\section{Introduction}
\label{sec:introduction}
Broad emission lines are prominent features in the ultraviolet and optical spectra of active galactic nuclei (AGNs) and quasars. These lines serve as powerful diagnostics of the gas dynamics in the broad-line region (BLR), which lies within the gravitational potential of the central supermassive black hole (SMBH). As such, they offer a unique method for estimating SMBH masses \citep[e.g.,][]{Woltjer1959, Peterson2000, Peterson2004, Gaskell2009}. Assuming the BLR gas is virialized, the black hole mass can be estimated using the relation:
\begin{equation}
    M_\mathrm{BH} = f \frac{R_\mathrm{BLR}\Delta V^{2}}{G}
\end{equation}
where $f$ is the dimensionless projection (or virial) factor accounting for the geometry and kinematics of the BLR, $G$ is the gravitational constant, $\Delta V$ is the velocity dispersion of the BLR clouds (typically inferred from the Doppler broadening of the emission lines), and $R_\mathrm{BLR}$ is the characteristic radius of the BLR. 

With extensive monitoring campaigns over the past decades, empirical global relations between the BLR size and AGN luminosity ($R_\mathrm{BLR} \propto L^{\beta}$) have been established. These so-called $R_\mathrm{BLR}$–$L$ relations hold for both low-ionization lines, such as $\textrm{Mg\,\textsc{ii}}$ ($\beta \simeq 0.7$; \citealt{Shen2016, Homayouni2020}) and the Balmer lines H$\alpha$/H$\beta$ ($\beta \simeq 0.5$; \citealt{Kaspi2005, Bentz2009, Bentz2013}), as well as for high-ionization lines like $\textrm{C\,\textsc{iv}}$  ($\beta \simeq 0.5$; \citealt{Lira2018, Kaspi2021}). The relatively tight intrinsic scatter of these relations suggests that the BLR radius can be inferred from a single-epoch measurement, thus providing a practical alternative to reverberation mapping for estimating black hole masses. This single-epoch virial method has become a fundamental tool for estimating black hole masses in large samples of AGNs and has been widely employed in studies exploring the co-evolution of SMBHs and their host galaxies.

Examining the variability of emission line widths with continuum or line luminosity—commonly referred to as the “breathing effect”—in AGNs with multi-epoch spectroscopic observations is a crucial approach for assessing the validity and estimating potential systematic uncertainties of single-epoch virial black hole mass estimates. However, the literature reports conflicting results regarding the breathing behavior.

In nearby Seyfert galaxies, reverberation mapping campaigns have shown that the H$\beta$ line width tends to decrease as luminosity increases \citep[e.g.,][]{Park2012, Barth2015}. Combined with the finding that the H$\beta$ lag scales with the mean continuum luminosity \citep[e.g.,][]{Peterson2002}, these results support a scenario in which the Balmer line-emitting region expands in response to enhanced ionizing radiation, consistent with photoionization model predictions \citep[e.g.,][]{Korista2004}. Similar breathing behavior in H$\beta$ has also been reported for large samples of quasars on average \citep{Shen2013, Wang2020}.

Nevertheless, other studies have reported a lack of breathing in either H$\beta$ or H$\alpha$ in quasars \citep[e.g.,][]{Guo2014,Ren2024II}.  Notably, while the SDSS Reverberation Mapping (SDSS-RM, \citealt{Shen2015}) sample analyzed by \citet{Wang2020} shows a clear average breathing trend in H$\beta$, many individual quasars within the sample exhibit either no breathing or even anti-breathing (i.e., line width increasing with luminosity). Moreover, their study reported no significant average breathing in H$\alpha$, in contrast to the behavior seen in H$\beta$.

The picture becomes more complex when considering other emission lines. For $\textrm{Mg\,\textsc{ii}}$, no or only weak breathing has been observed in quasars, including those with extreme variability \citep[e.g.,][]{Shen2013, Homan2020, Yang2020, Wang2020, Ren2024II}. In contrast, $\textrm{C\,\textsc{iv}}$  commonly exhibits anti-breathing behavior in quasars \citep[e.g.,][]{Richards2002, Wilhite2006, Shen2008, Richards2011}, likely reflecting the influence of non-virial kinematics such as outflows or inflows in the emitting region. However, a lack of breathing in $\textrm{C\,\textsc{iv}}$  has also been reported \citep{Ren2024II}.

Taken together, these results highlight the diversity and complexity of breathing behaviors across different emission lines and AGN populations. The observed variability patterns—ranging from normal breathing to anti-breathing and non-detections—not only reflect differences in the physical and kinematic conditions of the line-emitting regions but also suggest that the underlying mechanisms driving line width variability may differ across sources and lines. A more comprehensive understanding of breathing behaviors is thus essential for refining these mass estimates and for improving their applicability across cosmic time and AGN types.

Recently, \cite{Ren2024II} reported the absence of the breathing effect in H$\beta$, $\textrm{Mg\,\textsc{ii}}$, and $\textrm{C\,\textsc{iv}}$  for samples of extremely variable quasars, in contrast to previous studies of normal quasars. Since their data probe variability over much longer timescales ($\sim$3000 days in the observed frame), they speculated that the breathing effect, if present, may weaken or even disappear on such long timescales. This raises an intriguing question: is the breathing effect of different broad emission lines in regular quasars dependent on the timescale of variability?

In principle, the response of the broad line width to continuum variations is expected to be timescale dependent. On timescales comparable to the broad-line lag, any intrinsic breathing signal would be smeared out or significantly suppressed due to the finite light-travel time across the broad-line region (BLR). Notably, correcting for this time lag is unfeasible for most studies, as it requires dense reverberation mapping data and reliable lag measurements \citep[e.g.,][]{Wang2020}.

At slightly longer timescales, the impact of the BLR lag may diminish, potentially allowing the intrinsic breathing trend to emerge more clearly. On much longer timescales, however, additional physical processes, such as structural changes in the BLR or turbulence within the accretion disk \citep{Kang2021}, may begin to dominate, possibly altering or masking the breathing behavior.

In this work, we utilize an unprecedentedly large sample of quasars with multi-epoch spectroscopic observations from SDSS DR16 to systematically investigate the breathing effect in different broad emission lines. In particular, we present the first systematic study of the timescale dependence of quasar breathing by analyzing how the line width–luminosity correlation evolves across a wide range of temporal baselines.

The structure of this paper is as follows. In Sect.~\ref{sec:data reduction}, we describe the dataset and the spectral decomposition techniques employed to measure the properties of the broad emission lines. In Sect.~\ref{sec:breathing effect }, we examine the overall breathing effects for four broad emission lines in our sample. In Sect.~\ref{sec:timescale}, we introduce a novel method to divide the sample according to different time baselines and assess how the observed breathing behavior varies with the timescale of variability. 
In Sect.~\ref{sec:discussion}, we discuss our results and the implications of our findings. We conclude with a summary of our main findings in Sect.~\ref{sec:conclusions}.

Throughout this paper, we adopt a flat $\Lambda$CDM cosmology with $\Omega_{\Lambda} = 0.7$, $\Omega_\mathrm{m} = 0.3$, and $H_0 = 70\ \mathrm{km\ s^{-1}\ Mpc^{-1}}$.

\section{Data and reduction}
\label{sec:data reduction}
\subsection{Data}
To construct our sample, we required quasars to have at least two spectroscopic observations in order to enable variability studies. Starting from the full catalog of 750,414 quasars in SDSS Data Release 16 \citep[DR16Q,][]{DR16Q}, we performed a cross-match with the spectroscopic database using a 2$\arcsec$ search radius around each quasar's coordinates.

To ensure data consistency, we imposed additional criteria on the retrieved spectra: the relative difference between the pipeline redshifts of the two epochs must be less than 0.01 ($|\Delta z| / (z_{0}+1) < 0.01$, where $z_{0}$ is the primary redshift), removing irrelevant foreground/background galaxies.

\subsection{Spectral measurement}
\label{sec:spectral measurement}
We fit the retrieved spectra using the {\sc PyQSOFIT} code \citep{PyQSOFIT}, following the methodology described in \citet{Ren2024II}. The continuum was modeled as a power-law ($f_{\lambda} \propto \lambda^{\alpha}$) combined with a broadened $\textrm{Fe\,\textsc{ii}}$ emission template, fit over a set of carefully selected line-free continuum windows. After subtracting the best-fit continuum and $\textrm{Fe\,\textsc{ii}}$ emission, we modeled the residual broad and narrow emission lines using multiple Gaussian components to capture their complex profiles. Narrow Gaussian lines were constrained to have FWHM (Full Width at Half Maximum) $<$ 1200 km\,s$^{-1}$, while broad Gaussian lines were required to have FWHM $>$ 1200 km\,s$^{-1}$. 

The H$\alpha$ and H$\beta$ lines were each modeled with one narrow Gaussian and three broad Gaussians to fully represent their profiles. Each line in the [$\textrm{O\,\textsc{iii}}$] $\lambda\lambda$4959, 5007 doublet was modeled with one narrow (core) Gaussian and one broad (wing) Gaussian. The [$\textrm{N\,\textsc{ii}}$] $\lambda\lambda$6548, 6583 and [$\textrm{S\,\textsc{ii}}$] $\lambda\lambda$6716, 6731 doublets were each modeled with one narrow Gaussian per line. The $\textrm{Mg\,\textsc{ii}}$ and the $\textrm{C\,\textsc{iv}}$\footnote{Note that modeling the C IV doublet either as a single blended line or as two individual components ($\lambda$1548, $\lambda$1550) with appropriate flux ratios yields consistent results.} lines were each fitted with one narrow and two broad Gaussians. To minimize parameter degeneracies and ensure physical consistency, the widths and velocity offsets of the narrow H$\beta$ and [$\textrm{O\,\textsc{iii}}$] core components were tied together. Similarly, those of the narrow H$\alpha$, [$\textrm{N\,\textsc{ii}}$] and [$\textrm{S\,\textsc{ii}}$] components were linked. The flux ratios of the three line doublets ([$\textrm{O\,\textsc{iii}}$], [$\textrm{N\,\textsc{ii}}$] and [$\textrm{S\,\textsc{ii}}$]) were left as free parameters, and we verified that the fitted ratios remained within physically reasonable ranges.

We estimated the statistical uncertainties of our measured parameters using a Monte Carlo approach. Specifically, we generated 200 mock spectra by adding Gaussian noise to the original spectrum, with the noise level determined by the flux errors provided in the pipeline. Each mock spectrum was then fit using the same procedures as the original data, and the scatter in the resulting parameters was used to quantify their uncertainties.

\subsection{The final samples}
After excluding unphysical fits and applying signal-to-noise ratio (S/N) thresholds (requiring S/N $>$ 3 for the continuum and S/N $>$ 3 for the emission lines), we identified four subsets of quasars with multiple SDSS spectroscopic observations and successful spectral fitting. This yielded 1945 quasars with 6900 spectra in the H$\alpha$ sample, 5159 quasars with 16565 spectra in the H$\beta$ sample, 19313 quasars with 60965 spectra in the $\textrm{Mg\,\textsc{ii}}$ sample, and 6763 quasars with 21675 spectra in the $\textrm{C\,\textsc{iv}}$ sample.

\begin{figure}
    \centering
    \includegraphics[trim={0cm 0cm 0cm 0cm}, clip, width=\linewidth]{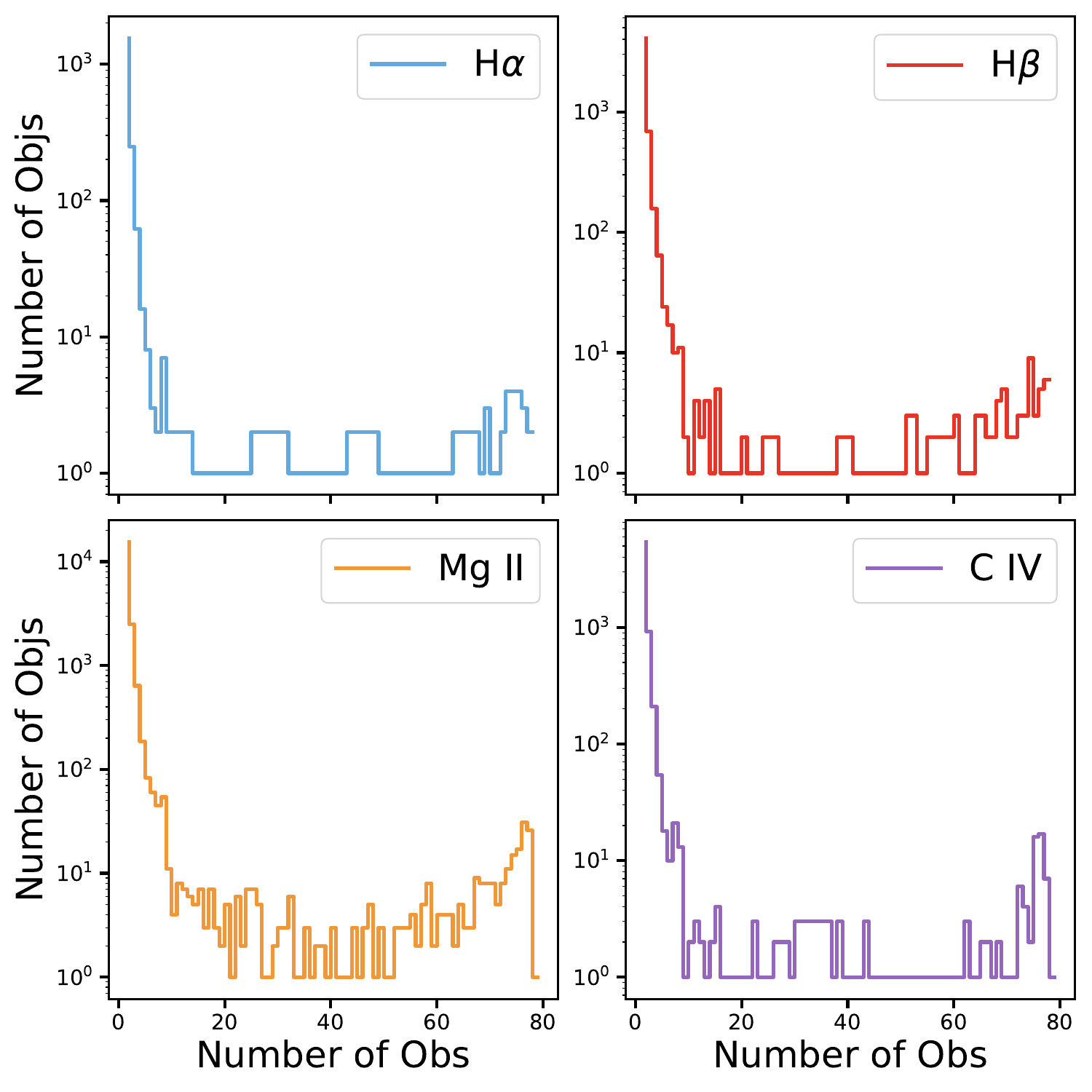}
    \caption{
    Distribution of sources in our samples by the number of repeated spectroscopic observations. The x-axis shows the number of observations per source, while the y-axis indicates the number of sources with that observation count. Only sources with more than one spectroscopic observation are shown, as those with a single observation are not included in our analysis.
    }
    \label{fig:obs_count}
\end{figure}

It is worth noting that some quasars appeared in multiple samples, as individual spectra could cover more than one broad emission line of interest. As illustrated in Fig.~\ref{fig:obs_count}, the majority of the quasars had only two spectroscopic observations, while only about 5\% of the sources have more than ten epochs. This observational cadence imposed a significant limitation on our ability to reliably measure broad-line time lags, which typically require a larger number of observations per object. Consequently, as is common in many previous studies, in this work we directly investigated the breathing behavior using multi-epoch spectroscopy without correcting for the effects of broad-line lags.

\section{Results}
\subsection{The overall breathing effect}
\label{sec:breathing effect }
\begin{figure}
    \centering
    \includegraphics[trim={0cm 0cm 0cm 0cm}, clip, width=\linewidth]{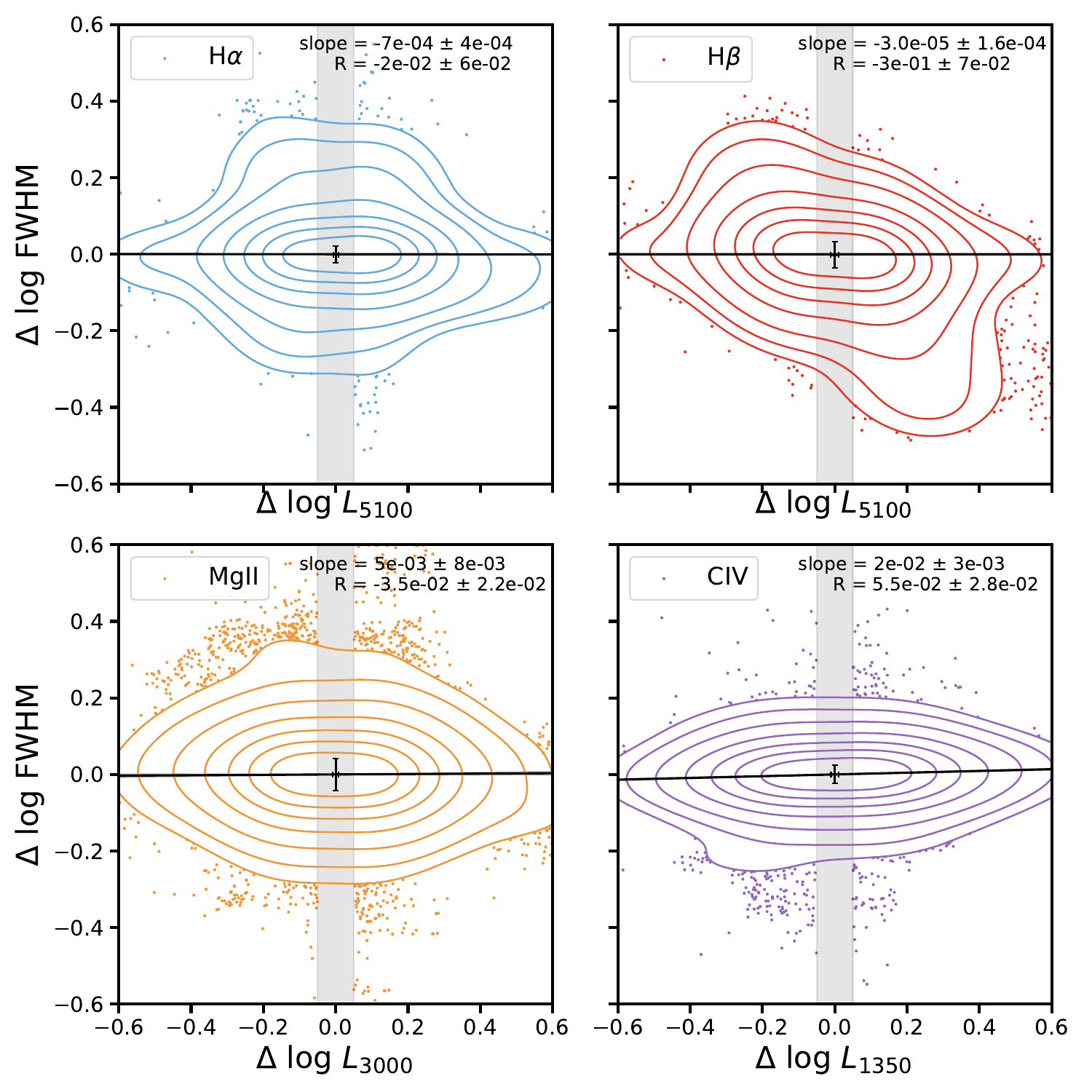}
    \caption{Variation in emission line FWHM as a function of the change in continuum luminosity, for H$\alpha$, H$\beta$, $\textrm{Mg\,\textsc{ii}}$, and $\textrm{C\,\textsc{iv}}$ , respectively. No significant breathing effect is detected in H$\alpha$, H$\beta$, or $\textrm{Mg\,\textsc{ii}}$, with regression slopes consistent with zero. A weak but statistically significant anti-breathing is observed in $\textrm{C\,\textsc{iv}}$ . In each panel, the black solid line denotes the best-fit regression slope (with the best-fit slope and correlation coefficient $R$ provided), and the error bar at the center represents the median statistical uncertainty of the data points. Interestingly, the H$\beta$ panel shows a clear excess of data points in the upper-left and lower-right regions, suggesting that strong breathing is present in a small subset of pairs, even though the sample as a whole shows no average breathing signal.
}
    \label{fig:breathing_effect}
\end{figure}

We first investigated the breathing behavior of the four broad emission lines in their respective samples, using the FWHM as a proxy for the velocity of the line emitting gas. The FWHM of each broad line profile was calculated from the total fitted profile obtained by summing over all broad Gaussian components. To quantify FWHM variability, we computed the difference in $\Delta (\log\mathrm{FWHM}$) between each epoch and all subsequent epochs for each quasar. For a source with $n$ spectroscopic observations, this yielded $n(n-1)/2$ observation pairs.

To characterize the relationship between FWHM and continuum variability, we fit a linear relation $y = \alpha x$ to all observation pairs, accounting for uncertainties in both $x$ and $y$ (denoted as $\Delta x$ and $\Delta y$). The best-fit slope $a$ was obtained by minimizing the following quantity:
\begin{equation} \chi^{2} = \sum_{i=1}^{N}\sum_{j=1}^{N_\mathrm{pairs}}\frac{w_{i}(y_{i,j}-\alpha x_{i,j})^{2}}{(\alpha \Delta x_{i,j})^{2}+(\Delta y_{i,j})^{2}}. \end{equation}  
where $w_i$ denoted the weight assigned to each data point. To avoid the fit being dominated by a small number of quasars with many observations, we assigned weights such that each quasar, rather than each observation pair, contributes equally to the regression (i.e., $w_i = 1/N_\mathrm{pairs}$ for each source).

This approach greatly increased the total number of data points. To robustly estimate the uncertainty in the regression slope, we applied a bootstrap resampling procedure to the parent sample, where sources were resampled as whole units, with all associated observation pairs included in each iteration. The linear slope was then computed for all the observation pairs within each resampled subset. The final slope was defined as the median slope of these resampled subsets, with the 1$\sigma$ confidence interval defined by the 16th and 84th percentiles. The Spearman $R$ value and its uncertainty were estimated in the same way.

To mitigate contamination from flux calibration uncertainties, we retained only data points where $\Delta \log L$ exceeds both 0.05 dex and 1.5 times its associated uncertainty, ensuring that the measured continuum changes are likely intrinsic. This filtering criterion was applied consistently in subsequent analyses. The regression slopes for H$\alpha$, H$\beta$ and $\textrm{Mg\,\textsc{ii}}$ lines are consistent with 0 (Fig. \ref{fig:breathing_effect}), indicating no significant evidence of breathing in these lines. In contrast, $\textrm{C\,\textsc{iv}}$  exhibits weak but statistically robust anti-breathing.  Interestingly, the H$\beta$ panel in Fig.~\ref{fig:breathing_effect} shows a clear excess of data points in the upper-left and lower-right regions, suggesting that strong breathing is present in a small subset of pairs, even though the sample as a whole shows no average breathing signal. The nature of these outliers is explored further in Sect.~\ref{sec:Hb}.

\subsection{The timescale dependency of the breathing effect}\label{sec:timescale}
To assess how the breathing effect  varies with timescale, we grouped all data points into bins based on different time intervals. For each pair of observations, we first calculated the rest-frame time interval as $\Delta t_\mathrm{rest} = \Delta t_\mathrm{obs} / (1 + z)$.
Given that the characteristic timescale of BLR breathing can vary significantly across quasars, it is natural to assume that this timescale is related to the size of the BLR. Therefore, when binning by timescale, we normalized the rest-frame time interval by the size of the BLR to account for source-to-source differences, yielding $\Delta t_\mathrm{scaled} = \Delta t_\mathrm{rest} / (R_\mathrm{BLR} / c)$, where $R_\mathrm{BLR}$ was estimated  using empirical scaling relations between BLR size and the quasar’s optical continuum luminosity.

For H$\alpha$ and H$\beta$, we adopted the $R_\mathrm{BLR}-L$ relation from \citet{Bentz2013}, given in Eq.~\ref{Eq1}, which uses the monochromatic luminosity at 5100\,\AA. Although H$\alpha$ and H$\beta$ are believed to originate from similar regions in the BLR, some studies suggest that the H$\alpha$ BLR may be slightly more extended \citep{Cho2023}. However, the effect of this difference is likely minor and remains uncertain, so we use the same relation for both lines:
\begin{equation}
    \label{Eq1}
    \log (R_\mathrm{BLR} / \mathrm{1\ lt}- \mathrm{day}) = 1.53\ +\ 0.53\ \log(\lambda L_{\lambda} / 10^{44} \mathrm{erg\ s^{-1}})
\end{equation}
For $\textrm{Mg\,\textsc{ii}}$, we adopted the scaling relation from \citet{Yu2023}, shown in Eq.~\ref{Eq2}, which uses the 3000\,\AA\ continuum luminosity. This relation is consistent with that of \citet{Homayouni2020} within uncertainties but exhibits significantly smaller intrinsic scatter, making it preferable for our analysis:
\begin{equation}
    \label{Eq2}
    \log (R_\mathrm{BLR} / \mathrm{1\ lt}- \mathrm{day}) = 2.07\ +\ 0.39\ \log(\lambda L_{\lambda} / 10^{45} \mathrm{erg\ s^{-1}})
\end{equation}
For $\textrm{C\,\textsc{iv}}$ , we used the relation from \citet{Grier2019}, based on the continuum luminosity at 1350\,\AA:
\begin{equation}
    \label{Eq3}
    \log (R_\mathrm{BLR} / \mathrm{1\ lt}- \mathrm{day}) = 0.92\ +\ 0.52\ \log (\lambda L_{\lambda} / 10^{44} \mathrm{erg\ s^{-1}})
\end{equation}

After normalizing the time intervals, we assigned weights ($w_{i}=1/N_\mathrm{pairs}$) to each data point such that each quasar, rather than each observation pair, contributes equally to the regression. We then performed linear fits to assess the breathing effect within each time-lag bin using the same methodology described above.

\begin{figure}
    \centering
    \includegraphics[trim={0cm 0cm 0cm 0cm}, clip, width=1.0\linewidth]{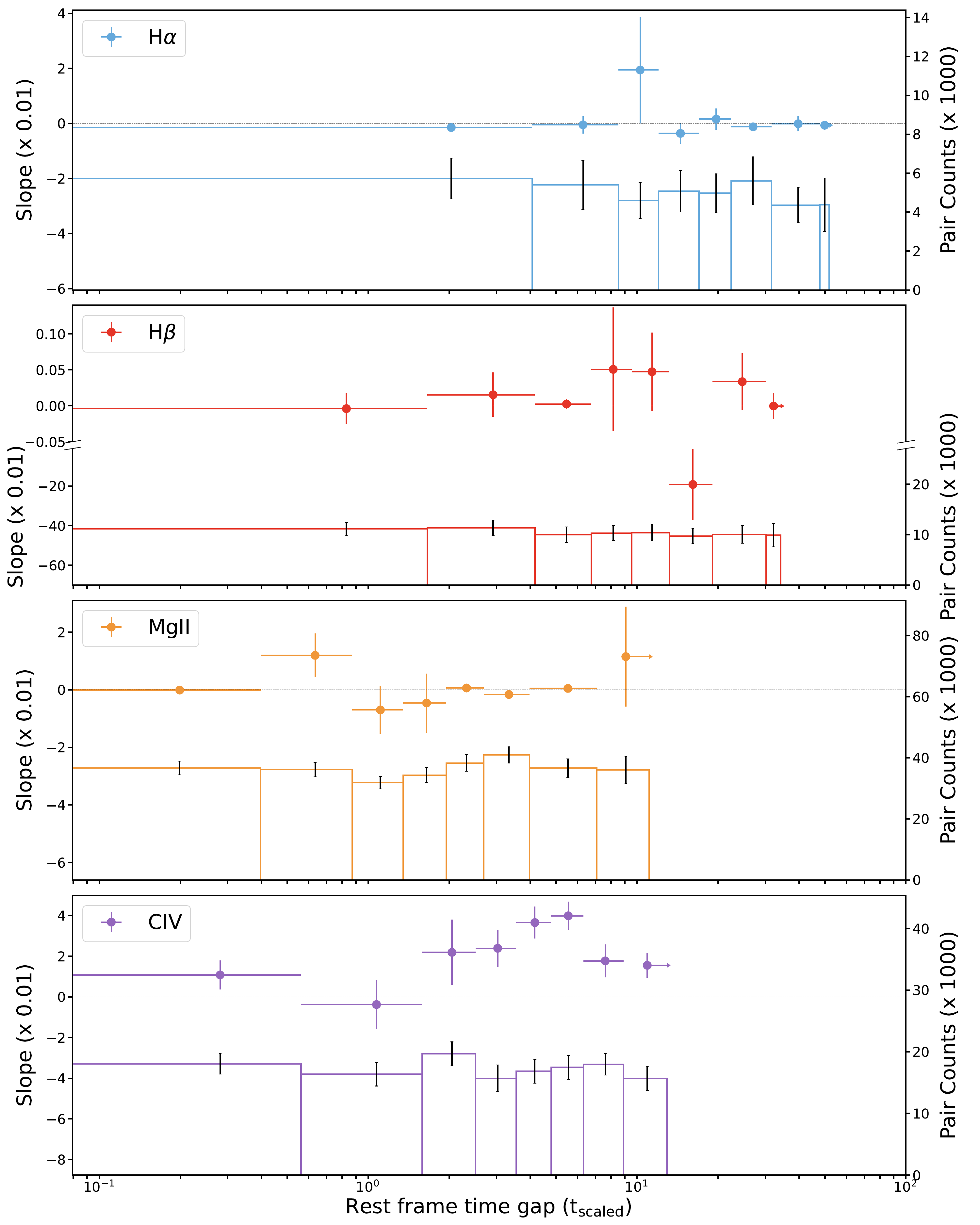}
    \caption{The breathing effect slope as a function of the rest-frame time interval between observations for all four broad emission lines. The bar plot shows the number of observation pairs in each bin, with black error bars indicating the scatter in the pair counts, estimated via bootstrapping.}
    \label{fig:all_line_BE}
\end{figure}

In Fig.~\ref{fig:all_line_BE}, we present how the breathing slope varies with $t_\mathrm{scaled}$ (from top to bottom: H$\alpha$, H$\beta$, $\textrm{Mg\,\textsc{ii}}$, and $\textrm{C\,\textsc{iv}}$ ). 

We find that the H$\alpha$, H$\beta$, and $\textrm{Mg\,\textsc{ii}}$ lines do not exhibit a clear breathing effect at any scaled timescale. While the absence of breathing at $t_{\mathrm{scaled}} < 1$ may be attributed to the impact of the broad-line region (BLR) lag, the continued lack of breathing at longer timescales suggests that these lines do not exhibit significant breathing behavior in our sample overall.

In contrast, the $\textrm{C\,\textsc{iv}}$  line exhibits a clear and timescale-dependent anti-breathing behavior at $t_{\mathrm{scaled}} > 1$. The anti-breathing trend becomes progressively stronger with increasing $t_{\mathrm{scaled}}$, peaks around $t_{\mathrm{scaled}} \sim 6$, and weakens at longer timescales.

We further examined whether the observed behaviors depend on continuum luminosity or redshift. To this end, we divided each parent sample into two bins based on continuum luminosity (or redshift), ensuring that each bin contains an approximately equal number of quasars. In the higher luminosity (or redshift) bin, the number of observation pairs in the longest time interval bin is consistently smaller, because of the limited temporal baseline. To mitigate the impact of this limitation, we adjust the selection of time interval bins accordingly. We then repeat the same analysis as performed on the full sample to derive the breathing effect  slope as a function of time interval for each subsample. We find no clear dependence of the breathing effect  slope on either continuum luminosity or redshift, for all the H$\alpha$, H$\beta$, $\textrm{Mg\,\textsc{ii}}$, and $\textrm{C\,\textsc{iv}}$  samples.

To better understand the line-specific behaviors, we next discuss the results for each emission line in detail.

\section{Discussion}\label{sec:discussion}
\subsection{H$\alpha$ and $\textrm{Mg\,\textsc{ii}}$}
While earlier studies have found little or no evidence of breathing in broad H$\alpha$ and $\textrm{Mg\,\textsc{ii}}$ lines \citep{Shen2013,Homan2020,Yang2020,Wang2020,Ren2024II}, our analysis extends this conclusion by showing that the absence of breathing holds across all examined timescales. This reinforces the view that, on average, these two low-ionization lines do not exhibit significant breathing behavior.

A plausible explanation for this lies in the physical structure of the broad-line region (BLR). Photoionization modeling by \citet{Guo2020} suggests that the H$\alpha$ and $\textrm{Mg\,\textsc{ii}}$ emissions may predominantly arise from gas located near the outer boundary of the BLR. In this scenario, variations in continuum luminosity would not appreciably alter the average radius of the line-emitting region. Consequently, the line widths would remain effectively unchanged, naturally leading to the absence of an observable breathing effect, consistent with our results.

It is important to emphasize, however, that the lack of a breathing effect in these low-ionization lines does not necessarily imply that the corresponding line-emitting gas is not virialized. A key assumption in using the breathing effect to test virial motion is that the intrinsic size-luminosity ($R$–$L$) relation follows the same form as the global relation. Yet, this assumption may not always hold. For instance, \citet{Peterson2002} found that in NGC~5548, the H$\beta$ lag scales with the mean continuum level as $\tau \propto F_{5100}^{0.95}$, significantly steeper than the global $R \propto L_{5100}^{0.5}$ relation derived from ensemble studies \citep{Bentz2013}.

In the specific case of H$\alpha$ and $\textrm{Mg\,\textsc{ii}}$, as argued by \citet{Guo2020}, the distance to the line-emitting gas may remain largely unchanged with continuum variations, leading to the absence of an intrinsic $R$–$L$ relation and, consequently, a lack of breathing effect, even if the gas is gravitationally bound and virialized.

\subsection{H$\beta$}
\label{sec:Hb}
\begin{figure}
    \centering
    \includegraphics[trim={0cm 0cm 0cm 0cm}, clip, width=1.0\linewidth]{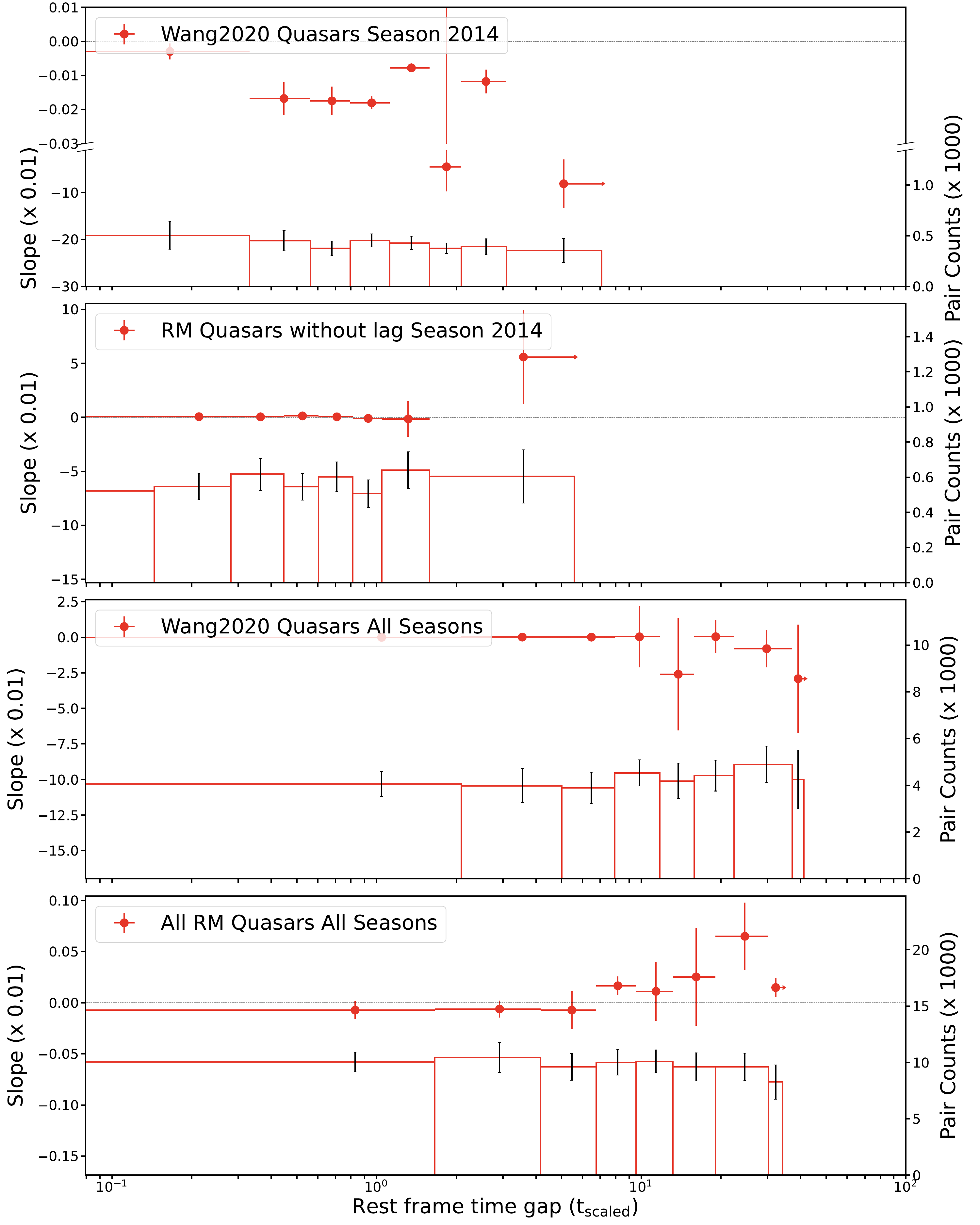}
    \caption{Same as the second panel of Fig.~\ref{fig:all_line_BE}, but for different SDSS-RM quasar subsamples.
    Top: 31 SDSS-RM quasars with well-measured H$\beta$ lags from \citet{Wang2020}, using only 2014 observations. A clear increase in the breathing slope is seen with increasing timescale.
    Middle: The same SDSS-RM quasar sample as above, but using all available SDSS-RM observations. The breathing trend disappears.
    Bottom: All RM quasars with more than 20 spectroscopic epochs. No breathing effect is detected either across all timescales.}
    \label{fig:Hb_RMAGN}
\end{figure}

Our analysis reveals a surprising absence of the breathing effect in the broad H$\beta$ line across all timescales (see the second panels of Fig.~\ref{fig:all_line_BE}). This is unexpected given that previous reverberation mapping (RM) studies have reported clear breathing behavior in some individual quasars \citep[e.g.,][]{Barth2015}, consistent with expectations from virialized broad-line region (BLR) motion \citep{Park2012}. \citet{Wang2020} further demonstrated that most quasars in a carefully selected subsample of SDSS RM quasars \citep{Shen2015} — those with reliably measured H$\beta$ lags \citep{Grier2017} — exhibit a significant breathing effect. Our lack of detection in the much broader sample is therefore particularly intriguing.

Closer inspection of the second panel of Fig.~\ref{fig:breathing_effect} reveals a distinct subset of data points in the upper-left and lower-right regions that exhibit strong breathing behavior. These points predominantly originate from SDSS-RM quasars, largely because this population contributes significantly more observation pairs and therefore dominates the sample. To better assess the underlying distribution, we calculated the breathing slope $\alpha$ and its statistical uncertainty $\alpha_{\mathrm{err}}$ for each pair, and identified breathing pairs using a stringent criterion of $\alpha < -3\alpha_{\mathrm{err}}$. Despite the difference in sample sizes, we find that both RM and non-RM quasars exhibit nearly identical fractions of normal breathing pairs (4.1\% vs. 4.3\%), suggesting that the RM quasars are not statistically distinct from the non-RM population in this regard. 

\begin{figure}
    \centering
    \includegraphics[trim={0cm 0cm 0cm 0cm}, clip, width=1.0\linewidth]{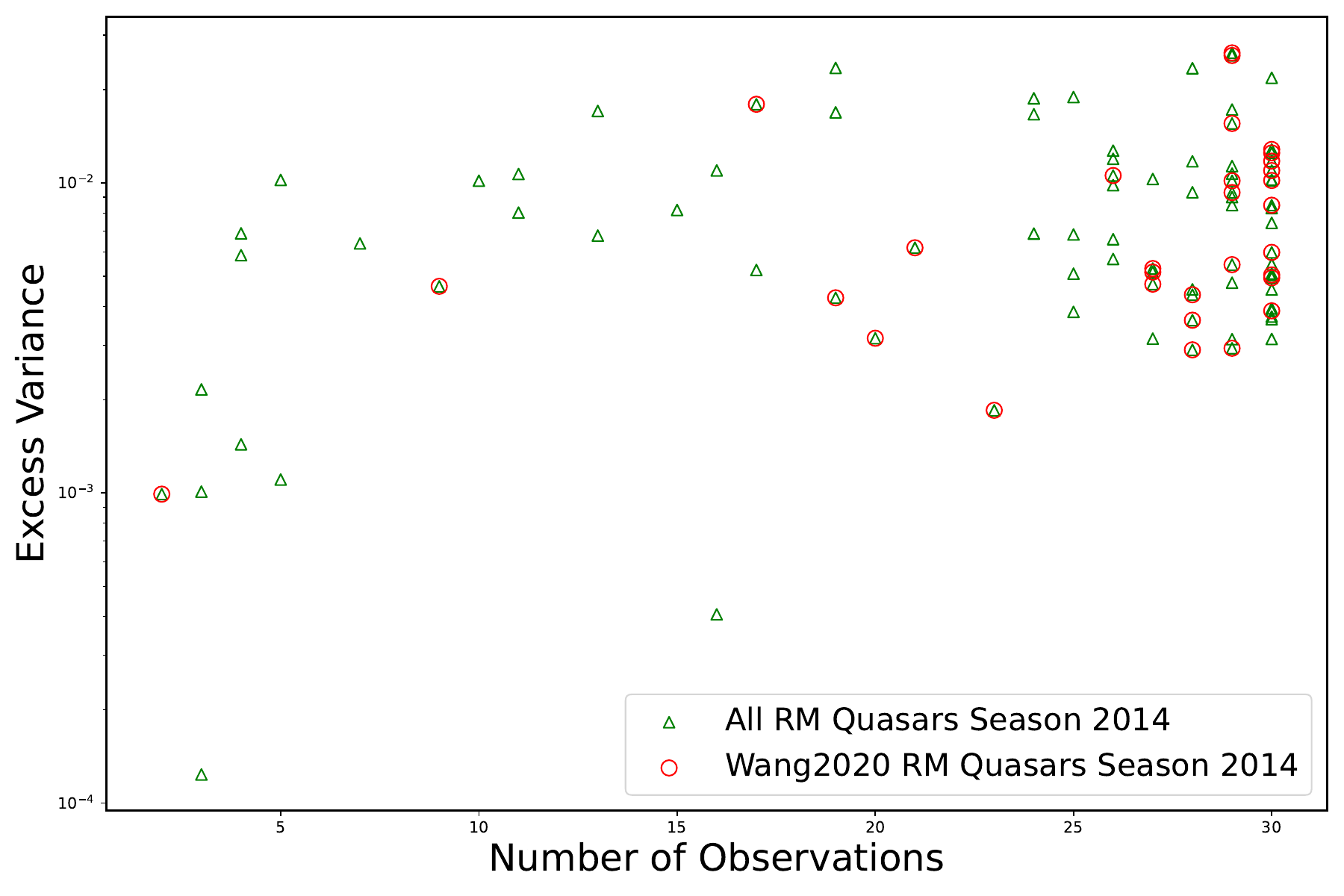}
    \caption{The x-axis shows the number of spectroscopic observations per SDSS RM quasar, and the y-axis shows the variation amplitude (excess variance). Red circles represent RM quasars with significant H$\beta$ lag detections from \citet{Wang2020}, while green triangles represent all RM quasars in our parent sample, including those without detectable lags. Both samples are based on data from the 2014 monitoring season. There is no clear distinction in either the number of observations or the excess variance between quasars with detected lags and those without.
 }
    \label{fig:excess_variance_vs_num_obs}
\end{figure}

To benchmark our method and results, we reanalyzed the H$\beta$ sample from \citet{Wang2020} without correcting for the effects of BLR lags. Following their approach, we focused exclusively on SDSS-RM spectroscopic observations from the season of 2014, during which reliable H$\beta$ lags were detected in these sources. The measured values were obtained using the fitting procedure described in Sect.~\ref{sec:spectral measurement}. As shown in the top panel of Fig.~\ref{fig:Hb_RMAGN}, this particular subsample exhibits a typical negative breathing slope across most time-interval bins. Notably, we observe a clear trend of increasing slope magnitude with longer timescales, reaching a value of $0.08 \pm 0.05$ at the longest timescale ($t_{\mathrm{scaled}} \gtrsim 5$), which is statistically consistent with the average slope of $-0.10 \pm 0.04$ reported by \citet{Wang2020}. This result supports the proposed scenario we introduced in Sect.~\ref{sec:introduction}: without correcting for BLR lag effects, the breathing signal can be smeared out at short timescales due to the time delay between the continuum and line emission. However, at longer timescales, the impact of lag smearing diminishes, allowing the underlying breathing effect to re-emerge. This further validates our method as a plausible and effective approach for identifying the breathing effect  when it is present.

However, when we extend the analysis of the same H$\beta$ quasar sample from \citet{Wang2020} to include all available epochs of SDSS-RM spectroscopic observations — not just those from the 2014 season — the breathing trend disappears (see the third panel of Fig.~\ref{fig:Hb_RMAGN}). This indicates that the presence of the breathing effect can vary with time, even for the same objects. Notably, the quasars selected by \citet{Wang2020} were chosen specifically because they exhibited detectable H$\beta$ BLR lags during the 2014 monitoring season, and they also showed clear breathing behavior during that period. This suggests a correlation between the detectability of H$\beta$ lags and the manifestation of the breathing effect.

We further note that the 31 H$\beta$ quasars analyzed in \citet{Wang2020} were drawn from a parent sample of 222 SDSS-RM quasars with H$\beta$ spectral coverage \citep{Grier2017}, meaning that the vast majority of SDSS-RM quasars do not have detectable H$\beta$ lags. To place these findings in a broader context, we analyzed all RM quasars in our parent sample. As shown in the bottom panel of Fig.~\ref{fig:Hb_RMAGN}, these sources likewise show no average H$\beta$ breathing signal. 

To further explore possible intrinsic differences between RM quasars with and without detected H$\beta$ lags, we compare their excess variance, which traces the intrinsic variability of the quasar, and the number of spectroscopic observations during the 2014 season. For each quasar, the excess variance $\sigma_\mathrm{rms}^2$ is calculated following \citet{Vaughan2003}:
\begin{equation}
\sigma_\mathrm{rms}^2 = \frac{1}{N-1} \sum (X_i - \bar{X})^2 - \frac{1}{N} \sum \sigma_i^2,
\end{equation}
where $N$ is the number of spectroscopic measurements, $X_i$ is the observed spectroscopic magnitude measured at 5100\AA, $\bar{X}$ is the average spectroscopic magnitude (converted from the continuum luminosity), and $\sigma_i$ is the uncertainty of each spectroscopic magnitude.
As shown in Fig.~\ref{fig:excess_variance_vs_num_obs}, there is no apparent difference between RM quasars with and without lag detections. Interestingly, those without detected H$\beta$ lags also show no evidence of H$\beta$ breathing during the 2014 season (see the second panel of Fig. \ref{fig:Hb_RMAGN}), further supporting a connection between the detectability of H$\beta$ lags and the manifestation of the H$\beta$ breathing effect. 

\begin{figure}
    \centering
    \includegraphics[trim={0cm 0cm 0cm 0cm}, clip, width=1.0\linewidth]{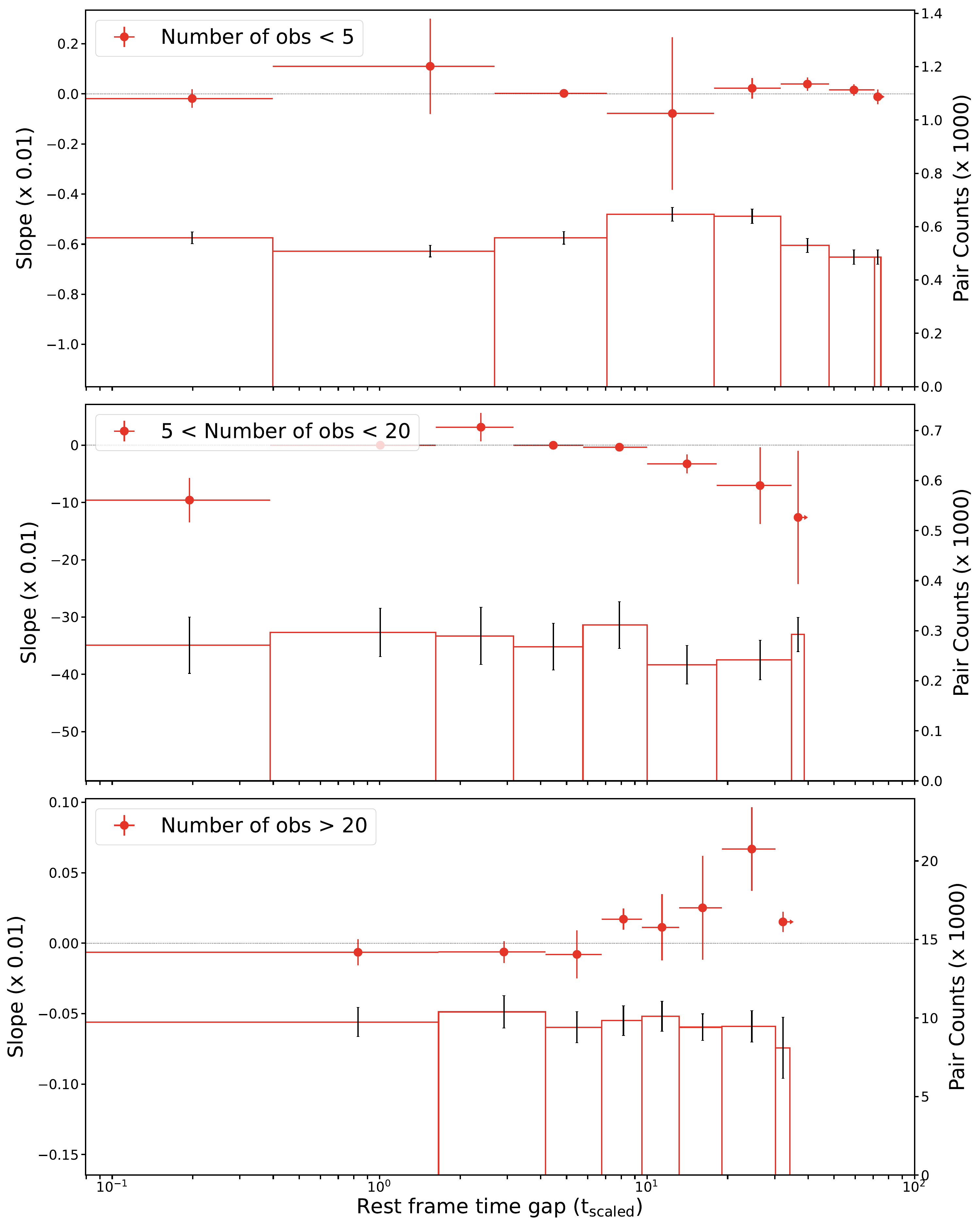}
    \caption{{Same as Fig.~\ref{fig:Hb_RMAGN} but for quasars with different number of spectroscopic observations. None of the subsamples exhibits significant breathing effect in H$\beta$.}}
    \label{fig:Hb_nobs}
\end{figure}

Taken together, we find no evidence for an average H$\beta$ breathing effect either among the entire quasar sample or within the full SDSS-RM quasar sample. The breathing effect is detected only in the 31 SDSS-RM quasars reported by \citet{Wang2020} that exhibit significant H$\beta$ lag detections, and only during the 2014 season when the lags were measured. We further examine whether the non-detection of H$\beta$ breathing might depend on the number of spectroscopic epochs per quasar; as shown in Fig.~\ref{fig:Hb_nobs}, no obvious trend is apparent.
These results suggest that while the H$\beta$ breathing effect can indeed manifest when a reliable BLR lag is detected, it is likely not a persistent or universal phenomenon across all epochs or quasars. The detectability of the breathing effect therefore could be be episodic. Future monitoring with longer duration and higher cadence could help to further test and confirm these findings.

This conclusion is further supported by our finding that only $\sim$4\% of all observation pairs exhibit a normal breathing signature, indicating that for the majority of the time, the H$\beta$ emission line does not show breathing behavior. In fact, a lack of breathing — or even anti-breathing — has been previously reported in individual objects \citep[e.g.,][]{Wang2020} and in different epochs of the same source \citep[e.g.,][]{Feng2024}. On longer timescales, \citet{Ren2024II} similarly found no significant breathing effect in a sample of extremely variable quasars (EVQs), consistent with our results.

A plausible explanation for the above findings is that, while the broad emission lines respond to variations in the ionizing continuum, the observed optical continuum variability — typically used as a proxy — does not always faithfully trace the ionizing flux. When the optical and ionizing continua are well correlated, both the BLR lag and the breathing effect can be detected. Conversely, if the two continua vary independently, neither phenomenon may be observable. This scenario could explain the lack of breathing detected in our analysis, and suggests that quasars may exhibit correlated optical and ionizing variability only during a small fraction of the time, leading to the episodic appearance of the breathing effect.

Indeed, anomalous responses of broad emission lines to continuum variability have been frequently reported \citep{Goad2016, Pei2017, Gaskell2021, Homayouni2024}. These findings are consistent with recent observational studies \citep{Xin2020, Sou2022} and physical models of quasar variability \citep{Cai2016, Cai2018, Cai2020}, which increasingly highlight that quasar variability across different bands—from optical to UV and X-ray—is not necessarily well correlated.

\subsection{$\textrm{C\,\textsc{iv}}$}
\label{sec:CIV}

\begin{figure}
    \centering
    \includegraphics[trim={0cm 0cm 0cm 0cm}, clip, width=1.0\linewidth]{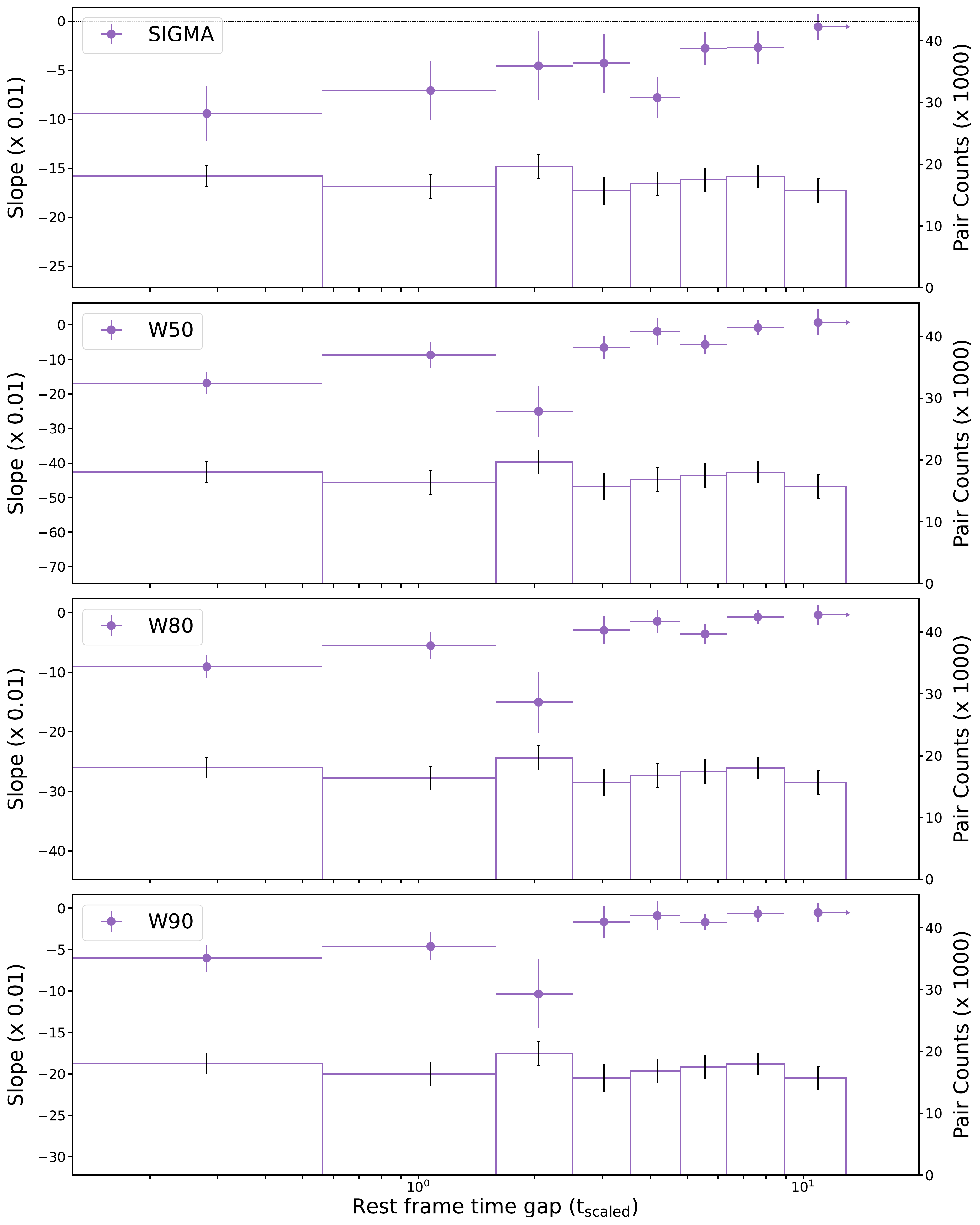}
    \caption{Same as the bottom panel of Fig.\ref{fig:all_line_BE}, but using alternative velocity-based line-width tracers to measure the breathing slopes of the $\textrm{C\,\textsc{iv}}$  line. We adopt line dispersion $\sigma$, as well as the non-parametric widths $w_{50}$, $w_{80}$, and $w_{90}$, which are more sensitive to the line wings and thus better trace the response of the broad, potentially reverberating component.}
    \label{fig:CIV_all_tracers}
\end{figure}

The anti-breathing trend of the $\textrm{C\,\textsc{iv}}$  line, as shown in Fig.~\ref{fig:breathing_effect}, is consistent with previous findings in the literature \citep{Richards2002, Wilhite2006, Shen2008, Richards2011, Wang2020}. A plausible interpretation for this behavior invokes a two-component origin of the $\textrm{C\,\textsc{iv}}$  emission: a broad, rapidly responding component associated with the classical broad-line region (BLR), and a narrower, less variable component likely originating from an intermediate line region or a radiatively driven disk wind. As the continuum luminosity increases, the relative contribution from the broad, reverberating component becomes more prominent, resulting in an overall increase in line width, manifesting as an anti-breathing behavior \citep[e.g.,][]{Wang2020}.

In Fig.~\ref{fig:all_line_BE}, we further show that this anti-breathing effect is timescale dependent: it strengthens with increasing scaled time separation ($t_{\mathrm{scaled}}$), peaks around $t_{\mathrm{scaled}} \sim 6$, and declines at longer timescales. This evolution can be naturally understood within the two-component framework. The broad, wing-dominated component requires time to fully respond to continuum variations, which amplifies the anti-breathing signal on intermediate timescales. At longer timescales, the effect weakens, either because the narrower component, initially less responsive, starts to vary with the continuum and dilutes the overall trend, or because the response of the broad component saturates or becomes modified by other processes such as structural evolution or changes in the disk wind launching condition. Extended spectroscopic monitoring will be critical to test these scenarios and to constrain the physical size and variability timescale of the narrow $\textrm{C\,\textsc{iv}}$ - emitting region.

To further probe the nature of $\textrm{C\,\textsc{iv}}$  breathing, we examined line dispersion ($\sigma$) and several velocity-based line-width tracers ($w_{50}$, $w_{80}$, and $w_{90}$, defined as the velocity differences between the 25th and 75th, 10th and 90th, and 5th and 95th percentiles of the cumulative flux, respectively). These tracers are more sensitive to the line wings than FWHM and are therefore better suited for tracing the broad, reverberating component. As shown in Fig.~\ref{fig:CIV_all_tracers}, all four tracers consistently exhibit a normal breathing pattern, i.e., line width decreases with increasing luminosity, in contrast to the anti-breathing behavior seen in FWHM. It is worth noting that using these alternative line-width tracers does not affect the results presented above for other emission lines.

The breathing signal in $\textrm{C\,\textsc{iv}}$  traced by these metrics is strongest at intermediate timescales. At longer timescales, the breathing signal of the wing component also weakens, consistent with the scenario that additional mechanisms, such as changes in the wind structure or the global BLR geometry, modifying the line profile beyond reverberation at longer timescales. These results also highlight the importance of timescale-dependent analyses, and motivate future long-term spectroscopic campaigns to fully characterize the dynamic nature of broad-line variability in quasars.

\subsection{Implication on the black hole mass measurement}
\label{sec:implication}

Different breathing effects across emission lines suggest that intrinsic quasar variability may introduce additional uncertainties in single-epoch black hole mass estimates. The scatter in single-epoch mass measurements is known to correlate with the continuum variability amplitude, and is commonly modeled as following $\delta \log M_\mathrm{BH} = (2\alpha + 0.5)\delta \log L$, where $\alpha$ is the breathing slope and $\delta \log L$ represents the variation amplitude of the continuum luminosity \citep{Shen2013}.

In this study, we detect no significant breathing effect  in the H$\alpha$, H$\beta$, and $\textrm{Mg\,\textsc{ii}}$ lines across a large quasar sample, while the $\textrm{C\,\textsc{iv}}$ line shows a timescale-dependent anti-breathing behavior. As we discussed earlier, the absence of a breathing signal does not necessarily imply that the corresponding line-emitting regions are not virialized. Nevertheless, these observational results have direct implications for the accuracy of single-epoch black hole mass estimates. In particular, AGNs with larger variability amplitudes are expected to suffer from larger mass uncertainties due to this effect. Moreover, given that quasar continuum variability at different wavelengths may be poorly correlated (see Sect.~\ref{sec:Hb}), the actual impact of $\delta \log L$ on mass estimates may be even more significant than what is inferred from the continuum variability amplitude measured near the wavelength of the emission line itself.

To quantify this effect empirically, we examine our quasars with exactly two spectroscopic observations and compute the black hole mass difference between the two epochs using the standard single-epoch virial mass formula from \citet{Shen2024}. We find that the median (90th percentile) mass differences are approximately 0.11 (0.22), 0.16 (0.30), and 0.10 (0.18) dex for H$\beta$-, $\textrm{Mg\textsc{ii}}$-, and $\textrm{C\textsc{iv}}$-based estimates, respectively. While these median values are relatively small compared to the typical systematic uncertainty of $\sim$0.4 dex, which is dominated by contributions from the virial factor uncertainty and $R$–$L$ relation scatter \citep[e.g.,][]{Shen2013}, a small fraction of quasars exhibit mass differences that are comparable to this level when measured across two spectroscopic epochs. Clearly, in the absence of strong breathing effects,  our results suggest that while continuum variability (and statistical fluctuations in line width measurements) contribute modestly to the typical scatter in single-epoch black hole mass estimates, they can introduce substantial deviations in a small fraction of quasars. 

\section{Conclusions}
\label{sec:conclusions}

In this work, we have performed a comprehensive analysis of the breathing effect—the luminosity-dependent variability of broad emission line widths—in a large sample of quasars with multi-epoch spectroscopy from SDSS DR16. By examining four prominent emission lines (H$\alpha$, H$\beta$, $\textrm{Mg\,\textsc{ii}}$, and $\textrm{C\,\textsc{iv}}$) and quantifying their line width response to continuum luminosity changes over a range of rest-frame time intervals, we aim to understand the physical conditions and temporal characteristics under which the breathing effect arises, and assess its implications for black hole mass estimates. Our main findings are summarized as follows:

\begin{enumerate}
\item 
No breathing effect on average is detected in the H$\alpha$, H$\beta$, or $\textrm{Mg\,\textsc{ii}}$ lines on any rest-frame timescale, suggesting that the widths of these low-ionization lines do not systematically respond to continuum luminosity changes in the majority of quasars.

\item
The $\textrm{C\,\textsc{iv}}$ line shows a clear anti-breathing effect, where the line width increases with continuum luminosity. This behavior is statistically significant and most pronounced at intermediate timescales, while at longer timescales, the trend may be influenced by additional physical processes that affect the $\textrm{C\,\textsc{iv}}$ profile.

\item
For H$\beta$, which has previously shown breathing in individual reverberation-mapped quasars, we find no significant trend on average. However, the breathing effect can be recovered in quasars with detected BLR lags, but only when restricting the analysis to epochs where the lag is clearly measurable. This suggests that the  breathing behaviors could be episodic, likely to be attributed to a weak or variable correlation between the optical and ionizing continua.

\item
Using quasars with two spectroscopic epochs, we find that black hole mass estimates typically differ by $\sim$0.1–0.2 dex, with $\sim$10\% of sources reaching differences of up to 0.3 dex. In the absence of a clear breathing effect, such differences cannot be easily corrected.
\end{enumerate}

These results offer new insights into the complicated nature of BLR breathing effect in quasars and reinforce the need for caution when using single-epoch spectra to estimate black hole masses.

\begin{acknowledgements} 
We thank the anonymous referee for the valuable comments and suggestions, which have significantly improved the manuscript.
This work was supported by the National Natural Science Foundation of China (grant Nos. 12533006, 12033006 \& 12192221) and the Cyrus Chun Ying Tang Foundations. 
\end{acknowledgements}

\bibliography{citation}{}

@ARTICLE{Woltjer1959,
       author = {{Woltjer}, L.},
        title = "{Emission Nuclei in Galaxies.}",
      journal = {\apj},
         year = 1959,
        month = jul,
       volume = {130},
        pages = {38},
          doi = {10.1086/146694},
       adsurl = {https://ui.adsabs.harvard.edu/abs/1959ApJ...130...38W}
}

@ARTICLE{Peterson2002,
       author = {{Peterson}, B.~M. and {Berlind}, P. and {Bertram}, R. and {Bischoff}, K. and {Bochkarev}, N.~G. and {Borisov}, N. and {Burenkov}, A.~N. and {Calkins}, M. and {Carrasco}, L. and {Chavushyan}, V.~H. and {Chornock}, R. and {Dietrich}, M. and {Doroshenko}, V.~T. and {Ezhkova}, O.~V. and {Filippenko}, A.~V. and {Gilbert}, A.~M. and {Huchra}, J.~P. and {Kollatschny}, W. and {Leonard}, D.~C. and {Li}, W. and {Lyuty}, V.~M. and {Malkov}, Yu. F. and {Matheson}, T. and {Merkulova}, N.~I. and {Mikhailov}, V.~P. and {Modjaz}, M. and {Onken}, C.~A. and {Pogge}, R.~W. and {Pronik}, V.~I. and {Qian}, B. and {Romano}, P. and {Sergeev}, S.~G. and {Sergeeva}, E.~A. and {Shapovalova}, A.~I. and {Spiridonova}, O.~I. and {Tao}, J. and {Tokarz}, S. and {Valdes}, J.~R. and {Vlasiuk}, V.~V. and {Wagner}, R.~M. and {Wilkes}, B.~J.},
        title = "{Steps toward Determination of the Size and Structure of the Broad-Line Region in Active Galactic Nuclei. XVI. A 13 Year Study of Spectral Variability in NGC 5548}",
      journal = {\apj},
         year = 2002,
        month = dec,
       volume = {581},
       number = {1},
        pages = {197-204},
          doi = {10.1086/344197},
archivePrefix = {arXiv},
       eprint = {astro-ph/0208064},
 primaryClass = {astro-ph},
       adsurl = {https://ui.adsabs.harvard.edu/abs/2002ApJ...581..197P}
}

@ARTICLE{Peterson2000,
       author = {{Peterson}, Bradley M. and {Wandel}, Amri},
        title = "{Evidence for Supermassive Black Holes in Active Galactic Nuclei from Emission-Line Reverberation}",
      journal = {\apjl},
         year = 2000,
        month = sep,
       volume = {540},
       number = {1},
        pages = {L13-L16},
          doi = {10.1086/312862},
archivePrefix = {arXiv},
       eprint = {astro-ph/0007147},
 primaryClass = {astro-ph},
       adsurl = {https://ui.adsabs.harvard.edu/abs/2000ApJ...540L..13P}
}

@ARTICLE{Peterson2004,
       author = {{Peterson}, B.~M. and {Ferrarese}, L. and {Gilbert}, K.~M. and {Kaspi}, S. and {Malkan}, M.~A. and {Maoz}, D. and {Merritt}, D. and {Netzer}, H. and {Onken}, C.~A. and {Pogge}, R.~W. and {Vestergaard}, M. and {Wandel}, A.},
        title = "{Central Masses and Broad-Line Region Sizes of Active Galactic Nuclei. II. A Homogeneous Analysis of a Large Reverberation-Mapping Database}",
      journal = {\apj},
         year = 2004,
        month = oct,
       volume = {613},
       number = {2},
        pages = {682-699},
          doi = {10.1086/423269},
archivePrefix = {arXiv},
       eprint = {astro-ph/0407299},
 primaryClass = {astro-ph},
       adsurl = {https://ui.adsabs.harvard.edu/abs/2004ApJ...613..682P}
}

@ARTICLE{DR16Q,
       author = {{Lyke}, Brad W. and {Higley}, Alexandra N. and {McLane}, J.~N. and {Schurhammer}, Danielle P. and {Myers}, Adam D. and {Ross}, Ashley J. and {Dawson}, Kyle and {Chabanier}, Sol{\`e}ne and {Martini}, Paul and {Busca}, Nicol{\'a}s G. and {Mas des Bourboux}, H{\'e}lion du and {Salvato}, Mara and {Streblyanska}, Alina and {Zarrouk}, Pauline and {Burtin}, Etienne and {Anderson}, Scott F. and {Bautista}, Julian and {Bizyaev}, Dmitry and {Brandt}, W.~N. and {Brinkmann}, Jonathan and {Brownstein}, Joel R. and {Comparat}, Johan and {Green}, Paul and {de la Macorra}, Axel and {Mu{\~n}oz Guti{\'e}rrez}, Andrea and {Hou}, Jiamin and {Newman}, Jeffrey A. and {Palanque-Delabrouille}, Nathalie and {P{\^a}ris}, Isabelle and {Percival}, Will J. and {Petitjean}, Patrick and {Rich}, James and {Rossi}, Graziano and {Schneider}, Donald P. and {Smith}, Alexander and {Vivek}, M. and {Weaver}, Benjamin Alan},
        title = "{The Sloan Digital Sky Survey Quasar Catalog: Sixteenth Data Release}",
      journal = {\apjs},
         year = 2020,
        month = sep,
       volume = {250},
       number = {1},
          eid = {8},
        pages = {8},
          doi = {10.3847/1538-4365/aba623},
archivePrefix = {arXiv},
       eprint = {2007.09001},
 primaryClass = {astro-ph.GA},
       adsurl = {https://ui.adsabs.harvard.edu/abs/2020ApJS..250....8L}
}

@ARTICLE{Shen2015,
       author = {{Shen}, Yue and {Brandt}, W.~N. and {Dawson}, Kyle S. and {Hall}, Patrick B. and {McGreer}, Ian D. and {Anderson}, Scott F. and {Chen}, Yuguang and {Denney}, Kelly D. and {Eftekharzadeh}, Sarah and {Fan}, Xiaohui and {Gao}, Yang and {Green}, Paul J. and {Greene}, Jenny E. and {Ho}, Luis C. and {Horne}, Keith and {Jiang}, Linhua and {Kelly}, Brandon C. and {Kinemuchi}, Karen and {Kochanek}, Christopher S. and {P{\^a}ris}, Isabelle and {Peters}, Christina M. and {Peterson}, Bradley M. and {Petitjean}, Patrick and {Ponder}, Kara and {Richards}, Gordon T. and {Schneider}, Donald P. and {Seth}, Anil and {Smith}, Robyn N. and {Strauss}, Michael A. and {Tao}, Charling and {Trump}, Jonathan R. and {Wood-Vasey}, W.~M. and {Zu}, Ying and {Eisenstein}, Daniel J. and {Pan}, Kaike and {Bizyaev}, Dmitry and {Malanushenko}, Viktor and {Malanushenko}, Elena and {Oravetz}, Daniel},
        title = "{The Sloan Digital Sky Survey Reverberation Mapping Project: Technical Overview}",
      journal = {\apjs},
         year = 2015,
        month = jan,
       volume = {216},
       number = {1},
          eid = {4},
        pages = {4},
          doi = {10.1088/0067-0049/216/1/4},
archivePrefix = {arXiv},
       eprint = {1408.5970},
 primaryClass = {astro-ph.IM},
       adsurl = {https://ui.adsabs.harvard.edu/abs/2015ApJS..216....4S}
}

@ARTICLE{Gaskell2021,
       author = {{Gaskell}, C. Martin and {Bartel}, Kayla and {Deffner}, Julia N. and {Xia}, Iris},
        title = "{Anomalous broad-line region responses to continuum variability in active galactic nuclei - I. H{\ensuremath{\beta}} variability}",
      journal = {\mnras},
         year = 2021,
        month = dec,
       volume = {508},
       number = {4},
        pages = {6077-6091},
          doi = {10.1093/mnras/stab2443},
archivePrefix = {arXiv},
       eprint = {1909.06275},
 primaryClass = {astro-ph.GA},
       adsurl = {https://ui.adsabs.harvard.edu/abs/2021MNRAS.508.6077G}
}

@ARTICLE{Sou2022,
       author = {{Sou}, Hao and {Wang}, Jun-Xian and {Xie}, Zhang-Liang and {Kang}, Wen-Yong and {Cai}, Zhen-Yi},
        title = "{The relation between X-ray and ultraviolet variability of quasars}",
      journal = {\mnras},
         year = 2022,
        month = jun,
       volume = {512},
       number = {4},
        pages = {5511-5519},
          doi = {10.1093/mnras/stac738},
archivePrefix = {arXiv},
       eprint = {2203.07119},
 primaryClass = {astro-ph.GA},
       adsurl = {https://ui.adsabs.harvard.edu/abs/2022MNRAS.512.5511S}
}

@software{PyQSOFIT,
       author = {{Guo}, Hengxiao and {Shen}, Yue and {Wang}, Shu},
        title = "{PyQSOFit: Python code to fit the spectrum of quasars}",
 howpublished = {Astrophysics Source Code Library, record ascl:1809.008},
         year = 2018,
        month = sep,
          eid = {ascl:1809.008},
       adsurl = {https://ui.adsabs.harvard.edu/abs/2018ascl.soft09008G}
}

@ARTICLE{Cai2020,
       author = {{Cai}, Zhen-Yi and {Wang}, Jun-Xian and {Sun}, Mouyuan},
        title = "{EUCLIA. II. On the Puzzling Large UV to X-Ray Lags in Seyfert Galaxies}",
      journal = {\apj},
         year = 2020,
        month = mar,
       volume = {892},
       number = {1},
          eid = {63},
        pages = {63},
          doi = {10.3847/1538-4357/ab7991},
archivePrefix = {arXiv},
       eprint = {2002.11116},
 primaryClass = {astro-ph.HE},
       adsurl = {https://ui.adsabs.harvard.edu/abs/2020ApJ...892...63C}
}

@ARTICLE{Cai2016,
       author = {{Cai}, Zhen-Yi and {Wang}, Jun-Xian and {Gu}, Wei-Min and {Sun}, Yu-Han and {Wu}, Mao-Chun and {Huang}, Xing-Xing and {Chen}, Xiao-Yang},
        title = "{Simulating the Timescale-Dependent Color Variation in Quasars with a Revised Inhomogeneous Disk Model}",
      journal = {\apj},
         year = 2016,
        month = jul,
       volume = {826},
       number = {1},
          eid = {7},
        pages = {7},
          doi = {10.3847/0004-637X/826/1/7},
archivePrefix = {arXiv},
       eprint = {1605.03185},
 primaryClass = {astro-ph.GA},
       adsurl = {https://ui.adsabs.harvard.edu/abs/2016ApJ...826....7C}
}

@ARTICLE{Cai2018,
       author = {{Cai}, Zhen-Yi and {Wang}, Jun-Xian and {Zhu}, Fei-Fan and {Sun}, Mou-Yuan and {Gu}, Wei-Min and {Cao}, Xin-Wu and {Yuan}, Feng},
        title = "{EUCLIA{\textemdash}Exploring the UV/Optical Continuum Lag in Active Galactic Nuclei. I. A Model without Light Echoing}",
      journal = {\apj},
         year = 2018,
        month = mar,
       volume = {855},
       number = {2},
          eid = {117},
        pages = {117},
          doi = {10.3847/1538-4357/aab091},
archivePrefix = {arXiv},
       eprint = {1711.06266},
 primaryClass = {astro-ph.HE},
       adsurl = {https://ui.adsabs.harvard.edu/abs/2018ApJ...855..117C}
}

@ARTICLE{Ren2024II,
       author = {{Ren}, Wenke and {Wang}, Junxian and {Cai}, Zhenyi and {Hu}, Xufan},
        title = "{Extreme Variability Quasars in Their Various States. II. Spectral Variation Revealed with Multiepoch Spectra}",
      journal = {\apj},
         year = 2024,
        month = mar,
       volume = {963},
       number = {1},
          eid = {7},
        pages = {7},
          doi = {10.3847/1538-4357/ad17cb},
archivePrefix = {arXiv},
       eprint = {2312.10869},
 primaryClass = {astro-ph.GA},
       adsurl = {https://ui.adsabs.harvard.edu/abs/2024ApJ...963....7R}
}

@ARTICLE{Grier2017,
       author = {{Grier}, C.~J. and {Trump}, J.~R. and {Shen}, Yue and {Horne}, Keith and {Kinemuchi}, Karen and {McGreer}, Ian D. and {Starkey}, D.~A. and {Brandt}, W.~N. and {Hall}, P.~B. and {Kochanek}, C.~S. and {Chen}, Yuguang and {Denney}, K.~D. and {Greene}, Jenny E. and {Ho}, L.~C. and {Homayouni}, Y. and {I-Hsiu Li}, Jennifer and {Pei}, Liuyi and {Peterson}, B.~M. and {Petitjean}, P. and {Schneider}, D.~P. and {Sun}, Mouyuan and {AlSayyad}, Yusura and {Bizyaev}, Dmitry and {Brinkmann}, Jonathan and {Brownstein}, Joel R. and {Bundy}, Kevin and {Dawson}, K.~S. and {Eftekharzadeh}, Sarah and {Fernandez-Trincado}, J.~G. and {Gao}, Yang and {Hutchinson}, Timothy A. and {Jia}, Siyao and {Jiang}, Linhua and {Oravetz}, Daniel and {Pan}, Kaike and {Paris}, Isabelle and {Ponder}, Kara A. and {Peters}, Christina and {Rogerson}, Jesse and {Simmons}, Audrey and {Smith}, Robyn and {Wang}, Ran},
        title = "{The Sloan Digital Sky Survey Reverberation Mapping Project: H{\ensuremath{\alpha}} and H{\ensuremath{\beta}} Reverberation Measurements from First-year Spectroscopy and Photometry}",
      journal = {\apj},
         year = 2017,
        month = dec,
       volume = {851},
       number = {1},
          eid = {21},
        pages = {21},
          doi = {10.3847/1538-4357/aa98dc},
archivePrefix = {arXiv},
       eprint = {1711.03114},
 primaryClass = {astro-ph.GA},
       adsurl = {https://ui.adsabs.harvard.edu/abs/2017ApJ...851...21G}
}

@ARTICLE{Bentz2013,
       author = {{Bentz}, Misty C. and {Denney}, Kelly D. and {Grier}, Catherine J. and {Barth}, Aaron J. and {Peterson}, Bradley M. and {Vestergaard}, Marianne and {Bennert}, Vardha N. and {Canalizo}, Gabriela and {De Rosa}, Gisella and {Filippenko}, Alexei V. and {Gates}, Elinor L. and {Greene}, Jenny E. and {Li}, Weidong and {Malkan}, Matthew A. and {Pogge}, Richard W. and {Stern}, Daniel and {Treu}, Tommaso and {Woo}, Jong-Hak},
        title = "{The Low-luminosity End of the Radius-Luminosity Relationship for Active Galactic Nuclei}",
      journal = {\apj},
         year = 2013,
        month = apr,
       volume = {767},
       number = {2},
          eid = {149},
        pages = {149},
          doi = {10.1088/0004-637X/767/2/149},
archivePrefix = {arXiv},
       eprint = {1303.1742},
 primaryClass = {astro-ph.CO},
       adsurl = {https://ui.adsabs.harvard.edu/abs/2013ApJ...767..149B}
}

@ARTICLE{Wang2020,
       author = {{Wang}, Shu and {Shen}, Yue and {Jiang}, Linhua and {Grier}, C.~J. and {Horne}, Keith and {Homayouni}, Y. and {Peterson}, B.~M. and {Trump}, Jonathan R. and {Brandt}, W.~N. and {Hall}, P.~B. and {Ho}, Luis C. and {Li}, Jennifer I. -Hsiu and {Hernandez Santisteban}, J.~V. and {Kinemuchi}, K. and {McGreer}, Ian D. and {Schneider}, D.~P.},
        title = "{The Sloan Digital Sky Survey Reverberation Mapping Project: How Broad Emission Line Widths Change When Luminosity Changes}",
      journal = {\apj},
         year = 2020,
        month = nov,
       volume = {903},
       number = {1},
          eid = {51},
        pages = {51},
          doi = {10.3847/1538-4357/abb36d},
archivePrefix = {arXiv},
       eprint = {2006.06178},
 primaryClass = {astro-ph.GA},
       adsurl = {https://ui.adsabs.harvard.edu/abs/2020ApJ...903...51W}
}

@ARTICLE{Park2012,
       author = {{Park}, Daeseong and {Woo}, Jong-Hak and {Treu}, Tommaso and {Barth}, Aaron J. and {Bentz}, Misty C. and {Bennert}, Vardha N. and {Canalizo}, Gabriela and {Filippenko}, Alexei V. and {Gates}, Elinor and {Greene}, Jenny E. and {Malkan}, Matthew A. and {Walsh}, Jonelle},
        title = "{The Lick AGN Monitoring Project: Recalibrating Single-epoch Virial Black Hole Mass Estimates}",
      journal = {\apj},
         year = 2012,
        month = mar,
       volume = {747},
       number = {1},
          eid = {30},
        pages = {30},
          doi = {10.1088/0004-637X/747/1/30},
archivePrefix = {arXiv},
       eprint = {1111.6604},
 primaryClass = {astro-ph.CO},
       adsurl = {https://ui.adsabs.harvard.edu/abs/2012ApJ...747...30P}
}

@ARTICLE{Xin2020,
       author = {{Xin}, Chengcheng and {Charisi}, Maria and {Haiman}, Zolt{\'a}n and {Schiminovich}, David},
        title = "{Correlation between optical and UV variability of a large sample of quasars}",
      journal = {\mnras},
         year = 2020,
        month = jun,
       volume = {495},
       number = {1},
        pages = {1403-1413},
          doi = {10.1093/mnras/staa1258},
archivePrefix = {arXiv},
       eprint = {2001.03154},
 primaryClass = {astro-ph.GA},
       adsurl = {https://ui.adsabs.harvard.edu/abs/2020MNRAS.495.1403X}
}

@ARTICLE{Barth2015,
       author = {{Barth}, Aaron J. and {Bennert}, Vardha N. and {Canalizo}, Gabriela and {Filippenko}, Alexei V. and {Gates}, Elinor L. and {Greene}, Jenny E. and {Li}, Weidong and {Malkan}, Matthew A. and {Pancoast}, Anna and {Sand}, David J. and {Stern}, Daniel and {Treu}, Tommaso and {Woo}, Jong-Hak and {Assef}, Roberto J. and {Bae}, Hyun-Jin and {Brewer}, Brendon J. and {Cenko}, S. Bradley and {Clubb}, Kelsey I. and {Cooper}, Michael C. and {Diamond-Stanic}, Aleksandar M. and {Hiner}, Kyle D. and {H{\"o}nig}, Sebastian F. and {Hsiao}, Eric and {Kandrashoff}, Michael T. and {Lazarova}, Mariana S. and {Nierenberg}, A.~M. and {Rex}, Jacob and {Silverman}, Jeffrey M. and {Tollerud}, Erik J. and {Walsh}, Jonelle L.},
        title = "{The Lick AGN Monitoring Project 2011: Spectroscopic Campaign and Emission-line Light Curves}",
      journal = {\apjs},
         year = 2015,
        month = apr,
       volume = {217},
       number = {2},
          eid = {26},
        pages = {26},
          doi = {10.1088/0067-0049/217/2/26},
archivePrefix = {arXiv},
       eprint = {1503.01146},
 primaryClass = {astro-ph.GA},
       adsurl = {https://ui.adsabs.harvard.edu/abs/2015ApJS..217...26B}
}

@ARTICLE{Yu2023,
       author = {{Yu}, Zhefu and {Martini}, Paul and {Penton}, A. and {Davis}, T.~M. and {Kochanek}, C.~S. and {Lewis}, G.~F. and {Lidman}, C. and {Malik}, U. and {Sharp}, R. and {Tucker}, B.~E. and {Aguena}, M. and {Annis}, J. and {Bertin}, E. and {Bocquet}, S. and {Brooks}, D. and {Carnero Rosell}, A. and {Carollo}, D. and {Carrasco Kind}, M. and {Carretero}, J. and {Costanzi}, M. and {da Costa}, L.~N. and {Pereira}, M.~E.~S. and {De Vicente}, J. and {Diehl}, H.~T. and {Doel}, P. and {Everett}, S. and {Ferrero}, I. and {Garc{\'\i}a-Bellido}, J. and {Gatti}, M. and {Gerdes}, D.~W. and {Gruen}, D. and {Gruendl}, R.~A. and {Gschwend}, J. and {Gutierrez}, G. and {Hinton}, S.~R. and {Hollowood}, D.~L. and {Honscheid}, K. and {James}, D.~J. and {Kuehn}, K. and {Mena-Fern{\'a}ndez}, J. and {Menanteau}, F. and {Miquel}, R. and {Nichol}, B. and {Paz-Chinch{\'o}n}, F. and {Pieres}, A. and {Plazas Malag{\'o}n}, A.~A. and {Raveri}, M. and {Romer}, A.~K. and {Sanchez}, E. and {Scarpine}, V. and {Sevilla-Noarbe}, I. and {Smith}, M. and {Suchyta}, E. and {Swanson}, M.~E.~C. and {Tarle}, G. and {Vincenzi}, M. and {Walker}, A.~R. and {Weaverdyck}, N.},
        title = "{OzDES Reverberation Mapping Programme: Mg II lags and R-L relation}",
      journal = {\mnras},
         year = 2023,
        month = jul,
       volume = {522},
       number = {3},
        pages = {4132-4147},
          doi = {10.1093/mnras/stad1224},
archivePrefix = {arXiv},
       eprint = {2208.05491},
 primaryClass = {astro-ph.GA},
       adsurl = {https://ui.adsabs.harvard.edu/abs/2023MNRAS.522.4132Y}
}

@ARTICLE{Homayouni2020,
       author = {{Homayouni}, Y. and {Trump}, Jonathan R. and {Grier}, C.~J. and {Horne}, Keith and {Shen}, Yue and {Brandt}, W.~N. and {Dawson}, Kyle S. and {Alvarez}, Gloria Fonseca and {Green}, Paul J. and {Hall}, P.~B. and {Hern{\'a}ndez Santisteban}, Juan V. and {Ho}, Luis C. and {Kinemuchi}, Karen and {Kochanek}, C.~S. and {Li}, Jennifer I. -Hsiu and {Peterson}, B.~M. and {Schneider}, D.~P. and {Starkey}, D.~A. and {Bizyaev}, Dmitry and {Pan}, Kaike and {Oravetz}, Daniel and {Simmons}, Audrey},
        title = "{The Sloan Digital Sky Survey Reverberation Mapping Project: Mg II Lag Results from Four Years of Monitoring}",
      journal = {\apj},
         year = 2020,
        month = sep,
       volume = {901},
       number = {1},
          eid = {55},
        pages = {55},
          doi = {10.3847/1538-4357/ababa9},
archivePrefix = {arXiv},
       eprint = {2005.03663},
 primaryClass = {astro-ph.GA},
       adsurl = {https://ui.adsabs.harvard.edu/abs/2020ApJ...901...55H}
}

@ARTICLE{Grier2019,
       author = {{Grier}, C.~J. and {Shen}, Yue and {Horne}, Keith and {Brandt}, W.~N. and {Trump}, J.~R. and {Hall}, P.~B. and {Kinemuchi}, K. and {Starkey}, David and {Schneider}, D.~P. and {Ho}, Luis C. and {Homayouni}, Y. and {I-Hsiu Li}, Jennifer and {McGreer}, Ian D. and {Peterson}, B.~M. and {Bizyaev}, Dmitry and {Chen}, Yuguang and {Dawson}, K.~S. and {Eftekharzadeh}, Sarah and {Guo}, Yucheng and {Jia}, Siyao and {Jiang}, Linhua and {Kneib}, Jean-Paul and {Li}, Feng and {Li}, Zefeng and {Nie}, Jundan and {Oravetz}, Audrey and {Oravetz}, Daniel and {Pan}, Kaike and {Petitjean}, Patrick and {Ponder}, Kara A. and {Rogerson}, Jesse and {Vivek}, M. and {Zhang}, Tianmeng and {Zou}, Hu},
        title = "{The Sloan Digital Sky Survey Reverberation Mapping Project: Initial C IV Lag Results from Four Years of Data}",
      journal = {\apj},
         year = 2019,
        month = dec,
       volume = {887},
       number = {1},
          eid = {38},
        pages = {38},
          doi = {10.3847/1538-4357/ab4ea5},
archivePrefix = {arXiv},
       eprint = {1904.03199},
 primaryClass = {astro-ph.GA},
       adsurl = {https://ui.adsabs.harvard.edu/abs/2019ApJ...887...38G}
}

@ARTICLE{Homan2020,
       author = {{Homan}, David and {MacLeod}, Chelsea L. and {Lawrence}, Andy and {Ross}, Nicholas P. and {Bruce}, Alastair},
        title = "{Behaviour of the Mg II 2798 {\r{A}} line over the full range of AGN variability}",
      journal = {\mnras},
         year = 2020,
        month = jul,
       volume = {496},
       number = {1},
        pages = {309-327},
          doi = {10.1093/mnras/staa1467},
archivePrefix = {arXiv},
       eprint = {1910.11364},
 primaryClass = {astro-ph.GA},
       adsurl = {https://ui.adsabs.harvard.edu/abs/2020MNRAS.496..309H}
}

@ARTICLE{Richards2002,
       author = {{Richards}, Gordon T. and {Vanden Berk}, Daniel E. and {Reichard}, Timothy A. and {Hall}, Patrick B. and {Schneider}, Donald P. and {SubbaRao}, Mark and {Thakar}, Anirudda R. and {York}, Donald G.},
        title = "{Broad Emission-Line Shifts in Quasars: An Orientation Measure for Radio-Quiet Quasars?}",
      journal = {\aj},
         year = 2002,
        month = jul,
       volume = {124},
       number = {1},
        pages = {1-17},
          doi = {10.1086/341167},
archivePrefix = {arXiv},
       eprint = {astro-ph/0204162},
 primaryClass = {astro-ph},
       adsurl = {https://ui.adsabs.harvard.edu/abs/2002AJ....124....1R}
}

@ARTICLE{Shen2024,
       author = {{Shen}, Yue and {Grier}, Catherine J. and {Horne}, Keith and {Stone}, Zachary and {Li}, Jennifer I. and {Yang}, Qian and {Homayouni}, Yasaman and {Trump}, Jonathan R. and {Anderson}, Scott F. and {Brandt}, W.~N. and {Hall}, Patrick B. and {Ho}, Luis C. and {Jiang}, Linhua and {Petitjean}, Patrick and {Schneider}, Donald P. and {Tao}, Charling and {Donnan}, Fergus. R. and {AlSayyad}, Yusra and {Bershady}, Matthew A. and {Blanton}, Michael R. and {Bizyaev}, Dmitry and {Bundy}, Kevin and {Chen}, Yuguang and {Davis}, Megan C. and {Dawson}, Kyle and {Fan}, Xiaohui and {Greene}, Jenny E. and {Gr{\"o}ller}, Hannes and {Guo}, Yucheng and {Ibarra-Medel}, H{\'e}ctor and {Jiang}, Yuanzhe and {Keenan}, Ryan P. and {Kollmeier}, Juna A. and {Lejoly}, Cassandra and {Li}, Zefeng and {de la Macorra}, Axel and {Moe}, Maxwell and {Nie}, Jundan and {Rossi}, Graziano and {Smith}, Paul S. and {Tee}, Wei Leong and {Weijmans}, Anne-Marie and {Xu}, Jiachuan and {Yue}, Minghao and {Zhou}, Xu and {Zhou}, Zhimin and {Zou}, Hu},
        title = "{The Sloan Digital Sky Survey Reverberation Mapping Project: Key Results}",
      journal = {\apjs},
         year = 2024,
        month = jun,
       volume = {272},
       number = {2},
          eid = {26},
        pages = {26},
          doi = {10.3847/1538-4365/ad3936},
archivePrefix = {arXiv},
       eprint = {2305.01014},
 primaryClass = {astro-ph.GA},
       adsurl = {https://ui.adsabs.harvard.edu/abs/2024ApJS..272...26S}
}

@ARTICLE{Kaspi2005,
       author = {{Kaspi}, Shai and {Maoz}, Dan and {Netzer}, Hagai and {Peterson}, Bradley M. and {Vestergaard}, Marianne and {Jannuzi}, Buell T.},
        title = "{The Relationship between Luminosity and Broad-Line Region Size in Active Galactic Nuclei}",
      journal = {\apj},
         year = 2005,
        month = aug,
       volume = {629},
       number = {1},
        pages = {61-71},
          doi = {10.1086/431275},
archivePrefix = {arXiv},
       eprint = {astro-ph/0504484},
 primaryClass = {astro-ph},
       adsurl = {https://ui.adsabs.harvard.edu/abs/2005ApJ...629...61K}
}

@ARTICLE{Bentz2009,
       author = {{Bentz}, Misty C. and {Peterson}, Bradley M. and {Netzer}, Hagai and {Pogge}, Richard W. and {Vestergaard}, Marianne},
        title = "{The Radius-Luminosity Relationship for Active Galactic Nuclei: The Effect of Host-Galaxy Starlight on Luminosity Measurements. II. The Full Sample of Reverberation-Mapped AGNs}",
      journal = {\apj},
         year = 2009,
        month = may,
       volume = {697},
       number = {1},
        pages = {160-181},
          doi = {10.1088/0004-637X/697/1/160},
archivePrefix = {arXiv},
       eprint = {0812.2283},
 primaryClass = {astro-ph},
       adsurl = {https://ui.adsabs.harvard.edu/abs/2009ApJ...697..160B}
}

@ARTICLE{Lira2018,
       author = {{Lira}, Paulina and {Kaspi}, Shai and {Netzer}, Hagai and {Botti}, Ismael and {Morrell}, Nidia and {Mej{\'\i}a-Restrepo}, Juli{\'a}n and {S{\'a}nchez-S{\'a}ez}, Paula and {Mart{\'\i}nez-Palomera}, Jorge and {L{\'o}pez}, Paula},
        title = "{Reverberation Mapping of Luminous Quasars at High z}",
      journal = {\apj},
         year = 2018,
        month = sep,
       volume = {865},
       number = {1},
          eid = {56},
        pages = {56},
          doi = {10.3847/1538-4357/aada45},
archivePrefix = {arXiv},
       eprint = {1806.08358},
 primaryClass = {astro-ph.GA},
       adsurl = {https://ui.adsabs.harvard.edu/abs/2018ApJ...865...56L}
}

@ARTICLE{Kaspi2021,
       author = {{Kaspi}, Shai and {Brandt}, W.~N. and {Maoz}, Dan and {Netzer}, Hagai and {Schneider}, Donald P. and {Shemmer}, Ohad and {Grier}, C.~J.},
        title = "{Taking a Long Look: A Two-decade Reverberation Mapping Study of High-luminosity Quasars}",
      journal = {\apj},
         year = 2021,
        month = jul,
       volume = {915},
       number = {2},
          eid = {129},
        pages = {129},
          doi = {10.3847/1538-4357/ac00aa},
       adsurl = {https://ui.adsabs.harvard.edu/abs/2021ApJ...915..129K}
}

@ARTICLE{Shen2016,
       author = {{Shen}, Yue and {Horne}, Keith and {Grier}, C.~J. and {Peterson}, Bradley M. and {Denney}, Kelly D. and {Trump}, Jonathan R. and {Sun}, Mouyuan and {Brandt}, W.~N. and {Kochanek}, Christopher S. and {Dawson}, Kyle S. and {Green}, Paul J. and {Greene}, Jenny E. and {Hall}, Patrick B. and {Ho}, Luis C. and {Jiang}, Linhua and {Kinemuchi}, Karen and {McGreer}, Ian D. and {Petitjean}, Patrick and {Richards}, Gordon T. and {Schneider}, Donald P. and {Strauss}, Michael A. and {Tao}, Charling and {Wood-Vasey}, W.~M. and {Zu}, Ying and {Pan}, Kaike and {Bizyaev}, Dmitry and {Ge}, Jian and {Oravetz}, Daniel and {Simmons}, Audrey},
        title = "{The Sloan Digital Sky Survey Reverberation Mapping Project: First Broad-line H{\ensuremath{\beta}} and Mg II Lags at z {\ensuremath{\gtrsim}}  0.3 from Six-month Spectroscopy}",
      journal = {\apj},
         year = 2016,
        month = feb,
       volume = {818},
       number = {1},
          eid = {30},
        pages = {30},
          doi = {10.3847/0004-637X/818/1/30},
archivePrefix = {arXiv},
       eprint = {1510.02802},
 primaryClass = {astro-ph.GA},
       adsurl = {https://ui.adsabs.harvard.edu/abs/2016ApJ...818...30S}
}

@ARTICLE{Korista2004,
       author = {{Korista}, Kirk T. and {Goad}, Michael R.},
        title = "{What the Optical Recombination Lines Can Tell Us about the Broad-Line Regions of Active Galactic Nuclei}",
      journal = {\apj},
         year = 2004,
        month = may,
       volume = {606},
       number = {2},
        pages = {749-762},
          doi = {10.1086/383193},
archivePrefix = {arXiv},
       eprint = {astro-ph/0402506},
 primaryClass = {astro-ph},
       adsurl = {https://ui.adsabs.harvard.edu/abs/2004ApJ...606..749K}
}

@ARTICLE{Yang2020,
       author = {{Yang}, Qian and {Shen}, Yue and {Chen}, Yu-Ching and {Liu}, Xin and {Annis}, James and {Avila}, Santiago and {Bertin}, Emmanuel and {Brooks}, David and {Buckley-Geer}, Elizabeth and {Carnero Rosell}, Aurelio and {Carrasco Kind}, Matias and {Carretero}, Jorge and {da Costa}, Luiz and {Desai}, Shantanu and {Thomas Diehl}, H. and {Doel}, Peter and {Frieman}, Josh and {Garcia-Bellido}, Juan and {Gaztanaga}, Enrique and {Gerdes}, David and {Gruen}, Daniel and {Gruendl}, Robert and {Gschwend}, Julia and {Gutierrez}, Gaston and {Hollowood}, Devon L. and {Honscheid}, Klaus and {Hoyle}, Ben and {James}, David and {Krause}, Elisabeth and {Kuehn}, Kyler and {Lidman}, Christopher and {Lima}, Marcos and {Maia}, Marcio and {Marshall}, Jennifer and {Martini}, Paul and {Menanteau}, Felipe and {Miquel}, Ramon and {Plazas Malag{\'o}n}, Andr{\'e}s and {Sanchez}, Eusebio and {Scarpine}, Vic and {Schindler}, Rafe and {Schubnell}, Michael and {Serrano}, Santiago and {Sevilla}, Ignacio and {Smith}, Mathew and {Soares-Santos}, Marcelle and {Sobreira}, Flavia and {Suchyta}, Eric and {Swanson}, Molly and {Tarle}, Gregory and {Vikram}, Vinu and {Walker}, Alistair},
        title = "{Spectral variability of a sample of extreme variability quasars and implications for the Mg II broad-line region}",
      journal = {\mnras},
         year = 2020,
        month = apr,
       volume = {493},
       number = {4},
        pages = {5773-5787},
          doi = {10.1093/mnras/staa645},
archivePrefix = {arXiv},
       eprint = {1904.10912},
 primaryClass = {astro-ph.GA},
       adsurl = {https://ui.adsabs.harvard.edu/abs/2020MNRAS.493.5773Y}
}

@ARTICLE{Shen2013,
       author = {{Shen}, Yue},
        title = "{The mass of quasars}",
      journal = {Bulletin of the Astronomical Society of India},
         year = 2013,
        month = mar,
       volume = {41},
       number = {1},
        pages = {61-115},
          doi = {10.48550/arXiv.1302.2643},
archivePrefix = {arXiv},
       eprint = {1302.2643},
 primaryClass = {astro-ph.CO},
       adsurl = {https://ui.adsabs.harvard.edu/abs/2013BASI...41...61S}
}

@ARTICLE{Guo2014,
       author = {{Guo}, Hengxiao and {Gu}, Minfeng},
        title = "{The Optical Variability of SDSS Quasars from Multi-epoch Spectroscopy. I. Results from 60 Quasars with >= Six-epoch Spectra}",
      journal = {\apj},
         year = 2014,
        month = sep,
       volume = {792},
       number = {1},
          eid = {33},
        pages = {33},
          doi = {10.1088/0004-637X/792/1/33},
archivePrefix = {arXiv},
       eprint = {1407.3025},
 primaryClass = {astro-ph.HE},
       adsurl = {https://ui.adsabs.harvard.edu/abs/2014ApJ...792...33G}
}

@ARTICLE{Guo2020,
       author = {{Guo}, Hengxiao and {Shen}, Yue and {He}, Zhicheng and {Wang}, Tinggui and {Liu}, Xin and {Wang}, Shu and {Sun}, Mouyuan and {Yang}, Qian and {Kong}, Minzhi and {Sheng}, Zhenfeng},
        title = "{Understanding Broad Mg II Variability in Quasars with Photoionization: Implications for Reverberation Mapping and Changing-look Quasars}",
      journal = {\apj},
         year = 2020,
        month = jan,
       volume = {888},
       number = {2},
          eid = {58},
        pages = {58},
          doi = {10.3847/1538-4357/ab5db0},
archivePrefix = {arXiv},
       eprint = {1907.06669},
 primaryClass = {astro-ph.GA},
       adsurl = {https://ui.adsabs.harvard.edu/abs/2020ApJ...888...58G}
}

@ARTICLE{Wilhite2006,
       author = {{Wilhite}, Brian C. and {Vanden Berk}, Daniel E. and {Brunner}, Robert J. and {Brinkmann}, Jonathan V.},
        title = "{Spectral Variability of Quasars in the Sloan Digital Sky Survey. II. The C IV Line}",
      journal = {\apj},
         year = 2006,
        month = apr,
       volume = {641},
       number = {1},
        pages = {78-89},
          doi = {10.1086/500421},
archivePrefix = {arXiv},
       eprint = {astro-ph/0512313},
 primaryClass = {astro-ph},
       adsurl = {https://ui.adsabs.harvard.edu/abs/2006ApJ...641...78W}
}

@ARTICLE{Shen2008,
       author = {{Shen}, Yue and {Greene}, Jenny E. and {Strauss}, Michael A. and {Richards}, Gordon T. and {Schneider}, Donald P.},
        title = "{Biases in Virial Black Hole Masses: An SDSS Perspective}",
      journal = {\apj},
         year = 2008,
        month = jun,
       volume = {680},
       number = {1},
        pages = {169-190},
          doi = {10.1086/587475},
archivePrefix = {arXiv},
       eprint = {0709.3098},
 primaryClass = {astro-ph},
       adsurl = {https://ui.adsabs.harvard.edu/abs/2008ApJ...680..169S}
}

@ARTICLE{Richards2011,
       author = {{Richards}, Gordon T. and {Kruczek}, Nicholas E. and {Gallagher}, S.~C. and {Hall}, Patrick B. and {Hewett}, Paul C. and {Leighly}, Karen M. and {Deo}, Rajesh P. and {Kratzer}, Rachael M. and {Shen}, Yue},
        title = "{Unification of Luminous Type 1 Quasars through C IV Emission}",
      journal = {\aj},
         year = 2011,
        month = may,
       volume = {141},
       number = {5},
          eid = {167},
        pages = {167},
          doi = {10.1088/0004-6256/141/5/167},
archivePrefix = {arXiv},
       eprint = {1011.2282},
 primaryClass = {astro-ph.GA},
       adsurl = {https://ui.adsabs.harvard.edu/abs/2011AJ....141..167R}
}

@ARTICLE{Gaskell2009,
       author = {{Gaskell}, C. Martin},
        title = "{What broad emission lines tell us about how active galactic nuclei work}",
      journal = {\nar},
         year = 2009,
        month = jul,
       volume = {53},
       number = {7-10},
        pages = {140-148},
          doi = {10.1016/j.newar.2009.09.006},
archivePrefix = {arXiv},
       eprint = {0908.0386},
 primaryClass = {astro-ph.CO},
       adsurl = {https://ui.adsabs.harvard.edu/abs/2009NewAR..53..140G}
}

@ARTICLE{Kang2021,
       author = {{Kang}, Wen-Yong and {Wang}, Jun-Xian and {Cai}, Zhen-Yi and {Ren}, Wen-Ke},
        title = "{More Variable Quasars Have Stronger Emission Lines}",
      journal = {\apj},
         year = 2021,
        month = apr,
       volume = {911},
       number = {2},
          eid = {148},
        pages = {148},
          doi = {10.3847/1538-4357/abeb69},
archivePrefix = {arXiv},
       eprint = {2103.01424},
 primaryClass = {astro-ph.GA},
       adsurl = {https://ui.adsabs.harvard.edu/abs/2021ApJ...911..148K}
}

@ARTICLE{Cho2023,
       author = {{Cho}, Hojin and {Woo}, Jong-Hak and {Wang}, Shu and {Son}, Donghoon and {Shin}, Jaejin and {Rakshit}, Suvendu and {Barth}, Aaron J. and {Bennert}, Vardha N. and {Gallo}, Elena and {Hodges-Kluck}, Edmund and {Treu}, Tommaso and {Bae}, Hyun-Jin and {Cho}, Wanjin and {Foord}, Adi and {Geum}, Jaehyuk and {Jadhav}, Yashashree and {Jeon}, Yiseul and {Kabasares}, Kyle M. and {Kang}, Daeun and {Kang}, Wonseok and {Kim}, Changseok and {Kim}, Donghwa and {Kim}, Minjin and {Kim}, Taewoo and {N. Le}, Huynh Anh and {Malkan}, Matthew A. and {Mandal}, Amit Kumar and {Park}, Daeseong and {Park}, Songyoun and {Sung}, Hyun-il and {U}, Vivian and {Williams}, Peter R.},
        title = "{The Seoul National University AGN Monitoring Project. IV. H{\ensuremath{\alpha}} Reverberation Mapping of Six AGNs and the H{\ensuremath{\alpha}} Size-Luminosity Relation}",
      journal = {\apj},
         year = 2023,
        month = aug,
       volume = {953},
       number = {2},
          eid = {142},
        pages = {142},
          doi = {10.3847/1538-4357/ace1e5},
archivePrefix = {arXiv},
       eprint = {2306.16683},
 primaryClass = {astro-ph.GA},
       adsurl = {https://ui.adsabs.harvard.edu/abs/2023ApJ...953..142C}
}

@ARTICLE{Feng2024,
       author = {{Feng}, Hai-Cheng and {Li}, Sha-Sha and {Bai}, J.~M. and {Liu}, H.~T. and {Lu}, Kai-Xing and {Pang}, Yu-Xuan and {Sun}, Mouyuan and {Wang}, Jian-Guo and {Zhang}, Yang-Wei and {Zhou}, Shuying},
        title = "{Velocity-resolved Reverberation Mapping of Changing-look Active Galactic Nucleus NGC 4151 during Outburst Stage. II. Results of Four Seasons of Observation}",
      journal = {\apj},
         year = 2024,
        month = dec,
       volume = {976},
       number = {2},
          eid = {176},
        pages = {176},
          doi = {10.3847/1538-4357/ad8568},
archivePrefix = {arXiv},
       eprint = {2409.01637},
 primaryClass = {astro-ph.GA},
       adsurl = {https://ui.adsabs.harvard.edu/abs/2024ApJ...976..176F}
}

@ARTICLE{Goad2016,
       author = {{Goad}, M.~R. and {Korista}, K.~T. and {De Rosa}, G. and {Kriss}, G.~A. and {Edelson}, R. and {Barth}, A.~J. and {Ferland}, G.~J. and {Kochanek}, C.~S. and {Netzer}, H. and {Peterson}, B.~M. and {Bentz}, M.~C. and {Bisogni}, S. and {Crenshaw}, D.~M. and {Denney}, K.~D. and {Ely}, J. and {Fausnaugh}, M.~M. and {Grier}, C.~J. and {Gupta}, A. and {Horne}, K.~D. and {Kaastra}, J. and {Pancoast}, A. and {Pei}, L. and {Pogge}, R.~W. and {Skielboe}, A. and {Starkey}, D. and {Vestergaard}, M. and {Zu}, Y. and {Anderson}, M.~D. and {Ar{\'e}valo}, P. and {Bazhaw}, C. and {Borman}, G.~A. and {Boroson}, T.~A. and {Bottorff}, M.~C. and {Brandt}, W.~N. and {Breeveld}, A.~A. and {Brewer}, B.~J. and {Cackett}, E.~M. and {Carini}, M.~T. and {Croxall}, K.~V. and {Dalla Bont{\`a}}, E. and {De Lorenzo-C{\'a}ceres}, A. and {Dietrich}, M. and {Efimova}, N.~V. and {Evans}, P.~A. and {Filippenko}, A.~V. and {Flatland}, K. and {Gehrels}, N. and {Geier}, S. and {Gelbord}, J.~M. and {Gonzalez}, L. and {Gorjian}, V. and {Grupe}, D. and {Hall}, P.~B. and {Hicks}, S. and {Horenstein}, D. and {Hutchison}, T. and {Im}, M. and {Jensen}, J.~J. and {Joner}, M.~D. and {Jones}, J. and {Kaspi}, S. and {Kelly}, B.~C. and {Kennea}, J.~A. and {Kim}, M. and {Kim}, S.~C. and {Klimanov}, S.~A. and {Lee}, J.~C. and {Leonard}, D.~C. and {Lira}, P. and {MacInnis}, F. and {Manne-Nicholas}, E.~R. and {Mathur}, S. and {McHardy}, I.~M. and {Montouri}, C. and {Musso}, R. and {Nazarov}, S.~V. and {Norris}, R.~P. and {Nousek}, J.~A. and {Okhmat}, D.~N. and {Papadakis}, I. and {Parks}, J.~R. and {Pott}, J. -U. and {Rafter}, S.~E. and {Rix}, H. -W. and {Saylor}, D.~A. and {Schimoia}, J.~S. and {Schn{\"u}lle}, K. and {Sergeev}, S.~G. and {Siegel}, M. and {Spencer}, M. and {Sung}, H. -I. and {Teems}, K.~G. and {Treu}, T. and {Turner}, C.~S. and {Uttley}, P. and {Villforth}, C. and {Weiss}, Y. and {Woo}, J. -H. and {Yan}, H. and {Young}, S. and {Zheng}, W. -K.},
        title = "{Space Telescope and Optical Reverberation Mapping Project. IV. Anomalous Behavior of the Broad Ultraviolet Emission Lines in NGC 5548}",
      journal = {\apj},
         year = 2016,
        month = jun,
       volume = {824},
       number = {1},
          eid = {11},
        pages = {11},
          doi = {10.3847/0004-637X/824/1/11},
archivePrefix = {arXiv},
       eprint = {1603.08741},
 primaryClass = {astro-ph.GA},
       adsurl = {https://ui.adsabs.harvard.edu/abs/2016ApJ...824...11G}
}

@ARTICLE{Pei2017,
       author = {{Pei}, L. and {Fausnaugh}, M.~M. and {Barth}, A.~J. and {Peterson}, B.~M. and {Bentz}, M.~C. and {De Rosa}, G. and {Denney}, K.~D. and {Goad}, M.~R. and {Kochanek}, C.~S. and {Korista}, K.~T. and {Kriss}, G.~A. and {Pogge}, R.~W. and {Bennert}, V.~N. and {Brotherton}, M. and {Clubb}, K.~I. and {Dalla Bont{\`a}}, E. and {Filippenko}, A.~V. and {Greene}, J.~E. and {Grier}, C.~J. and {Vestergaard}, M. and {Zheng}, W. and {Adams}, Scott M. and {Beatty}, Thomas G. and {Bigley}, A. and {Brown}, Jacob E. and {Brown}, Jonathan S. and {Canalizo}, G. and {Comerford}, J.~M. and {Coker}, Carl T. and {Corsini}, E.~M. and {Croft}, S. and {Croxall}, K.~V. and {Deason}, A.~J. and {Eracleous}, Michael and {Fox}, O.~D. and {Gates}, E.~L. and {Henderson}, C.~B. and {Holmbeck}, E. and {Holoien}, T.~W. -S. and {Jensen}, J.~J. and {Johnson}, C.~A. and {Kelly}, P.~L. and {Kim}, S. and {King}, A. and {Lau}, M.~W. and {Li}, Miao and {Lochhaas}, Cassandra and {Ma}, Zhiyuan and {Manne-Nicholas}, E.~R. and {Mauerhan}, J.~C. and {Malkan}, M.~A. and {McGurk}, R. and {Morelli}, L. and {Mosquera}, Ana and {Mudd}, Dale and {Muller Sanchez}, F. and {Nguyen}, M.~L. and {Ochner}, P. and {Ou-Yang}, B. and {Pancoast}, A. and {Penny}, Matthew T. and {Pizzella}, A. and {Poleski}, Rados{\l}aw and {Runnoe}, Jessie and {Scott}, B. and {Schimoia}, Jaderson S. and {Shappee}, B.~J. and {Shivvers}, I. and {Simonian}, Gregory V. and {Siviero}, A. and {Somers}, Garrett and {Stevens}, Daniel J. and {Strauss}, M.~A. and {Tayar}, Jamie and {Tejos}, N. and {Treu}, T. and {Van Saders}, J. and {Vican}, L. and {Villanueva}, Jr., S. and {Yuk}, H. and {Zakamska}, N.~L. and {Zhu}, W. and {Anderson}, M.~D. and {Ar{\'e}valo}, P. and {Bazhaw}, C. and {Bisogni}, S. and {Borman}, G.~A. and {Bottorff}, M.~C. and {Brandt}, W.~N. and {Breeveld}, A.~A. and {Cackett}, E.~M. and {Carini}, M.~T. and {Crenshaw}, D.~M. and {De Lorenzo-C{\'a}ceres}, A. and {Dietrich}, M. and {Edelson}, R. and {Efimova}, N.~V. and {Ely}, J. and {Evans}, P.~A. and {Ferland}, G.~J. and {Flatland}, K. and {Gehrels}, N. and {Geier}, S. and {Gelbord}, J.~M. and {Grupe}, D. and {Gupta}, A. and {Hall}, P.~B. and {Hicks}, S. and {Horenstein}, D. and {Horne}, Keith and {Hutchison}, T. and {Im}, M. and {Joner}, M.~D. and {Jones}, J. and {Kaastra}, J. and {Kaspi}, S. and {Kelly}, B.~C. and {Kennea}, J.~A. and {Kim}, M. and {Kim}, S.~C. and {Klimanov}, S.~A. and {Lee}, J.~C. and {Leonard}, D.~C. and {Lira}, P. and {MacInnis}, F. and {Mathur}, S. and {McHardy}, I.~M. and {Montouri}, C. and {Musso}, R. and {Nazarov}, S.~V. and {Netzer}, H. and {Norris}, R.~P. and {Nousek}, J.~A. and {Okhmat}, D.~N. and {Papadakis}, I. and {Parks}, J.~R. and {Pott}, J. -U. and {Rafter}, S.~E. and {Rix}, H. -W. and {Saylor}, D.~A. and {Schn{\"u}lle}, K. and {Sergeev}, S.~G. and {Siegel}, M. and {Skielboe}, A. and {Spencer}, M. and {Starkey}, D. and {Sung}, H. -I. and {Teems}, K.~G. and {Turner}, C.~S. and {Uttley}, P. and {Villforth}, C. and {Weiss}, Y. and {Woo}, J. -H. and {Yan}, H. and {Young}, S. and {Zu}, Y.},
        title = "{Space Telescope and Optical Reverberation Mapping Project. V. Optical Spectroscopic Campaign and Emission-line Analysis for NGC 5548}",
      journal = {\apj},
         year = 2017,
        month = mar,
       volume = {837},
       number = {2},
          eid = {131},
        pages = {131},
          doi = {10.3847/1538-4357/aa5eb1},
archivePrefix = {arXiv},
       eprint = {1702.01177},
 primaryClass = {astro-ph.GA},
       adsurl = {https://ui.adsabs.harvard.edu/abs/2017ApJ...837..131P}
}

@ARTICLE{Homayouni2024,
       author = {{Homayouni}, Y. and {Kriss}, Gerard A. and {De Rosa}, Gisella and {Plesha}, Rachel and {Cackett}, Edward M. and {Goad}, Michael R. and {Korista}, Kirk T. and {Horne}, Keith and {Fischer}, Travis and {Waters}, Tim and {Barth}, Aaron J. and {Kara}, Erin A. and {Landt}, Hermine and {Arav}, Nahum and {Boizelle}, Benjamin D. and {Bentz}, Misty C. and {Brotherton}, Michael S. and {Chelouche}, Doron and {Dalla Bont{\`a}}, Elena and {Dehghanian}, Maryam and {Du}, Pu and {Ferland}, Gary J. and {Fian}, Carina and {Gelbord}, Jonathan and {Grier}, Catherine J. and {Hall}, Patrick B. and {Hu}, Chen and {Ili{\'c}}, Dragana and {Joner}, Michael D. and {Kaastra}, Jelle and {Kaspi}, Shai and {Kova{\v{c}}evi{\'c}}, Andjelka B. and {Kynoch}, Daniel and {Li}, Yan-Rong and {Mehdipour}, Missagh and {Miller}, Jake A. and {Mitchell}, Jake and {Montano}, John and {Netzer}, Hagai and {Neustadt}, J.~M.~M. and {Partington}, Ethan and {Popovi{\'c}}, Luka {\v{C}}. and {Proga}, Daniel and {Storchi-Bergmann}, Thaisa and {Sanmartim}, David and {Siebert}, Matthew R. and {Treu}, Tommaso and {Vestergaard}, Marianne and {Wang}, Jian-Min and {Ward}, Martin J. and {Zaidouni}, Fatima and {Zu}, Ying},
        title = "{AGN STORM 2. V. Anomalous Behavior of the C IV Light Curve of Mrk 817}",
      journal = {\apj},
         year = 2024,
        month = mar,
       volume = {963},
       number = {2},
          eid = {123},
        pages = {123},
          doi = {10.3847/1538-4357/ad1be4},
archivePrefix = {arXiv},
       eprint = {2308.00742},
 primaryClass = {astro-ph.GA},
       adsurl = {https://ui.adsabs.harvard.edu/abs/2024ApJ...963..123H}
}

@ARTICLE{Vaughan2003,
       author = {{Vaughan}, S. and {Edelson}, R. and {Warwick}, R.~S. and {Uttley}, P.},
        title = "{On characterizing the variability properties of X-ray light curves from active galaxies}",
      journal = {\mnras},
         year = 2003,
        month = nov,
       volume = {345},
       number = {4},
        pages = {1271-1284},
          doi = {10.1046/j.1365-2966.2003.07042.x},
archivePrefix = {arXiv},
       eprint = {astro-ph/0307420},
 primaryClass = {astro-ph},
       adsurl = {https://ui.adsabs.harvard.edu/abs/2003MNRAS.345.1271V}
}
\bibliographystyle{aa}

\end{document}